\documentclass[prd,aps,graphics,twocolumn,nofootinbib]{revtex4}
\usepackage{amsmath,amssymb,amsfonts}
\usepackage{graphicx,epsfig,rotate,psfig}
\usepackage{dcolumn}
\usepackage{bm}
\usepackage{epsfig}
\def\dd{{\rm d}}
\def\eff{{\rm eff}}
\def\ppn{{\rm PPN}}
\def\scalefac{a}
\def\mat{{\rm m}}
\def\de{{\rm de}}
\def\rcto{{\rm c0}}
\def\rc{{\rm c}}

\def\ppn{\mathrm{PPN}}
\def\cav{\mathrm{cav}}
\def\etal{{\it et al.}}

\begin{document}

\title{The acceleration of the universe and the physics behind it}

\author{Jean--Philippe Uzan}
 \email{uzan@iap.fr}
 \affiliation{
             Institut d'Astrophysique de Paris,
             Universit\'e Pierre~\&~Marie Curie - Paris VI,
             CNRS-UMR 7095, 98 bis, Bd Arago, 75014 Paris, France.}

\date{\today}
\begin{abstract}
Using a general classification of dark enegy models in four classes, we discuss the
complementarity of cosmological observations to tackle down the physics beyond the acceleration of
our universe. We discuss the tests distinguishing the four classes and then focus on the dynamics
of the perturbations in the Newtonian regime. We also exhibit explicitely models that have
identical predictions for a subset of observations.
\end{abstract}
\pacs{{ \bf PACS numbers:} } \vskip2pc \maketitle \vskip 0.15cm
\section{Introduction}\label{sec-intro}

The flow of new data~\cite{wmap3,data1,data2,data3,data4,data5} has allowed to narrow the
constraints on the standard cosmological parameters. Among all the conclusions concerning our
universe, the existence of a recent acceleration phase seems to be more and more settled. Even
though, cosmology has a minimal standard model, consistent and robust with 6 or 7 free parameters
(the concordance model), many extensions (both of the primordial physics and of the matter
content) are still weakly constrained today. These parameters start to be measured very accurately
but it may turn out that our (successfull) parameterization may be too simple.

The quest for the understanding of the origin of this acceleration is however just starting (see
e.g. Refs.~\cite{ct,book,peebles03,pad1}). Different ways of attacking this problem have been
proposed~\cite{ct,bean,teg} using various properties to order the questions on the origin of the
acceleration. In this article, we come back to the classification proposed in
Refs.~\cite{uam,ctes2} and detail the degeneracies that have to be taken into account before
conclusions are drawn. This completes other attempts to define strategies to get insight into the
physics of dark energy\footnote{We emphasize here that the cause of the acceleration of the
universe is decoupled from the cosmological constant problem which is somehow assumed to be
solved. One goal of the dynamical dark energy models is to avoid the fine tuning related to the
cosmic coincidence problem.}.

The conclusion that our universe is accelerating assumes in the first place the validity of the
Copernician principle, that is the fact that the observable universe can be described on large
scales by a Friedmann-Lema\^{\i}tre spacetime so that its dynamics is characterized by a single
function of time, the scale factor $a(t)$. Since most data are related to events observed on our
past light cone, there is an intrinsic degeneracy along this cone. In the Friedmann-Lema\^{\i}tre
models, this degeneracy is lifted by the symmetry assumption. In the next to simple case, the
universe can be described by a spherically symmetric spacetime. For an observer seating at the
center of the universe, the degeneracy along the past-null cone entangles the cosmic time $t$ and
the radial distance $r$. At  low redshift, it can be described by a Lema\^{\i}tre-Tolman-Bondi
(LTB) spacetime~\cite{ltb} that depends on two arbitrary functions of $r$. Interestingly it was
shown that a LTB universe reproducing the luminosity distance-redshift relation observed can be
reconstructed without introducing any new form of matter~\cite{ltb1} and it was shown that
homogeneity cannot be proven using only background quantities~\cite{mus}. This implies that the
Copernician principle has to be tested observationally (see e.g. Refs.~\cite{obscosm}) as much as
possible, particularly under the light of some new proposals~\cite{kmr}.

We will not consider this interesting possibility in the following and we assume that the universe
is well described by a Friedmann-Lema\^{\i}tre spacetime. From a cosmological point of view, the
dynamics of the background expansion is characterized by the matter content of the universe, that
is a list of fluids\footnote{As long as the background dynamics is concerned, the symmetry imposed
by the cosmological principle is such that only perfect fluids can be considered. This is why the
parameterization of dark energy reduces to an equation of state at this level. Consider a comoving
observer with 4-velocity $u^\mu$ perpendicular to the hypersurfaces of homogeneity. The line
element of the FL spacetime takes the form $\dd s^2 = -(u_\mu \dd x^\mu)^2 + \gamma_{\mu\nu} \dd
x^\mu\dd x^\nu$ with $\gamma_{\mu\nu}=g_{\mu\nu}+u_\mu u_\nu$. As a symmetric rank-2 tensor, the
stress-energy tensor of any matter should decompose as $T_{\mu\nu}=A u_\mu u_\nu +
B\gamma_{\mu\nu}$ and $A=T_{\mu\nu}u^\mu u^\nu$ reduces to the energy density measured by this
comoving observer and $B$ to the isotropic pressure. Thus, there is no extra-assumption at this
stage.} with their equation of state. All background observations are then related to the function
$H(a)/H_0$ where $H_0$ is the Hubble constant today. The extra degrees of freedom, often referred
to as {\it dark energy} and needed to explain the data, can be introduced as a new kind of matter
or as a new property of gravity.

Let us clarify this. General relativity relies on two assumptions: (1) gravitational interaction
is described by a massless spin-2 field and (2) matter is minimally coupled to the metric which,
in particular, implies the validity of the weak equivalence principle. It follows that
\begin{equation}\label{ac1}
 S=\frac{c^3}{16\pi G}\int R\sqrt{-g}\dd^4x + S_m[{\rm mat};g_{\mu\nu}]
\end{equation}
where $g_{\mu\nu}$ is the metric tensor and $R$ its Ricci scalar. $S_m[{\rm mat};g_{\mu\nu}]$ is
the action of the matter fields. Each class of models will account for a modification of the
minimal action and our discussion restricts itself to models relying on a field theory. This has
the advantage to discuss which are the new degrees of freedom and to which extent the theory is
well defined. It also implies that we do not consider models in which the Friedmann equations are
modified without an underlying field theory which we consider on the same footing as an ad hoc
parameterization of the equation of state.

In the first approach one assumes that gravitation is described by general relativity while
introducing new forms of gravitating components beyond the standard model of particle physics to
explain the observed acceleration of the universe. This means that one adds a new term $S_{\rm
de}[\psi;g_{\mu\nu}]$ in the action~(\ref{ac1}) while keeping the Einstein-Hilbert action and the
coupling of all the fields (standard matter and dark matter) unchanged.

The other route is to allow for a modification of gravity. This means that the only long range
force that cannot be screened is assumed not to be described by general relativity.  Various ways
to extend the minimal action~(\ref{ac1}) have been considered, modifying the Einstein-Hilbert
action or the coupling of matter. Whatever the modification, one has to introduce new degrees of
freedom (not necessarily scalar) and even though we refer to these models as a modification of
gravity, we have to keep in mind that {\it they involve also new matter}.

\begin{figure}[htb]
 \centerline{\epsfig{figure=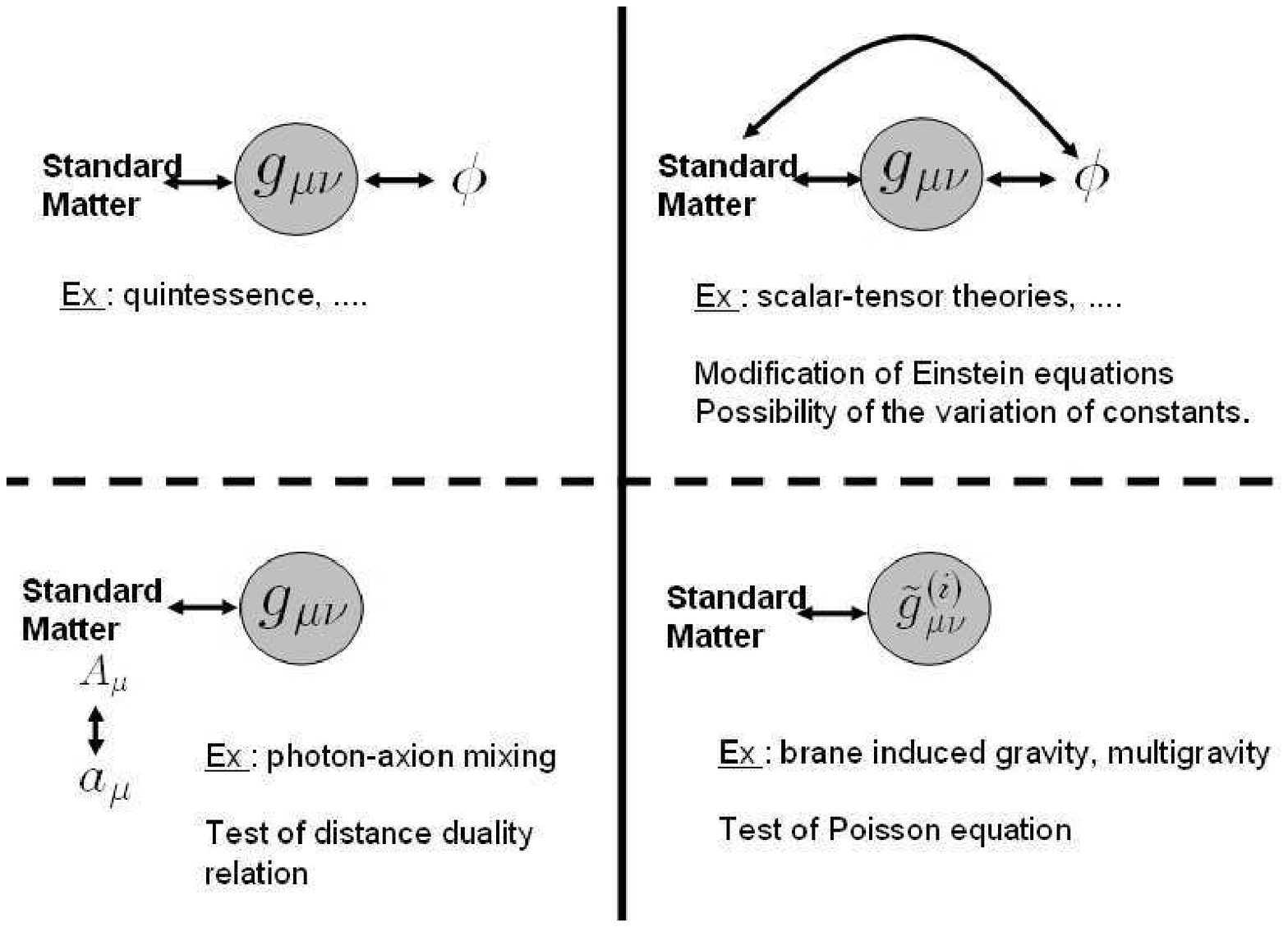,width=8cm}}
 \caption{Summary of the proposed different classes of models. As discussed in the text,
 various tests can be designed to distinguish between them. The classes differ according
 to the kind of new fields and to the way they couple to the metric
 $g_{\mu\nu}$ and to the standard matter fields. Left column accounts for models
 where gravitation is described by general relativity while right column models
 describe a modification of gravity. In the upper line classes, the new fields
 dominate the matter content of the universe at low redshift.
 Upper-left models (class A) consist of
 models in which a new kind of gravitating matter is introduced. In the upper-right
 models (class C), a light field induces a long-range force so that gravity is
 not described by a massless spin-2 graviton only. This is the case of
 scalar-tensor theories of gravity. In this class, Einstein equations
 are modified and there may be a variation of the fundamental
 constants. The lower-right models (class D) corresponds to models in which there
 may exist massive gravitons, such as in some class of braneworld
 scenarios. These models predict a modification of the Poisson
 equation on large scales. In the last models (lower-left, class B), the
 distance duality relation may be violated.}
 \label{fig0}
\end{figure}

This being defined, we can distinguish four general classes~\cite{uam}\footnote{A similar
classification can be applied to dark matter issue. The list of models cited here is indeed
incomplete and we are not aiming at an exhaustive review of specific models. See
Ref.~\cite{ct,book,peebles03} for that purpose.}. First are models in which gravity is not
modified
\begin{enumerate}
 \item[1.]\underline{Class A} consists of models in which the acceleration is
 driven by the gravitational effect of new fields which are minimally coupled to
 gravity. It follows that they are not coupled to the standard matter fields
 or to dark matter and that one is adding a new sector
 $$
 S_{\rm de}[{\rm de};g_{\mu\nu}]
 $$
 to the action~(\ref{ac1}). To explain the acceleration of the universe, the new matter must have
 an equation of state, $w=P/\rho$ smaller than $-1/3$.
 As standard examples of this class of models, we can
 cite the  standard {\it $\Lambda$CDM} model (which has the property to involve only 1 new parameter
 and no new field),
 {\it quintessence} models~\cite{quintessence} which invoke a canonical scalar
 field slow-rolling today, {\it solid dark matter} models~\cite{soliddm} induced by frustrated
 topological defects networks,
 {\it tachyon} models~\cite{tachyons}, {\it Chaplygin gaz} models~\cite{chap} which try to unify
 dark matter and dark energy, and {\it K-essence} models~\cite{kessence} models invoking scalar fields
 with a non-canonical
 kinetic term.
 \item[2.]\underline{Class B} introduces new fields which do not dominate the matter
 content so that they do not change the expansion rate of the universe. These
 fields are coupled to photons and thus affect the observed dynamics of
 the universe but not the dynamics itself. In particular they are not required to have an
 equation of state smaller than $-1/3$.
 An example is provided by {\it photon-axion oscillations} in an external magnetic field~\cite{mix}
 which aims at explaining the dimming of supernovae, not by an acceleration expansion but by the
 fact that part of the photons has oscillated into invisible axions. In that particular case, the
 electromagnetic sector is modified according to
 $$
 S_{\rm em}[A_\mu;g_{\mu\nu}] \rightarrow S_{\rm em}[A_\mu, a_\mu;g_{\mu\nu}].
 $$
 A specific signature of these models would be a violation of the distance duality
 relation~\cite{uam,bkaxion}, a possible violation of the variation of CMB temperature with
 redshift [$T_\gamma\propto(1+z)$] which seems to hold observationnally~\cite{petit} and in the
 future, the determination of the luminosity from gravitational waves will be insensitive to such
 a coupling.
\end{enumerate}

Then come models with a modification of gravity. Once such a possibility is considered, many new
models arise~\cite{will}. Indeed in considering these modifications, one needs to take great care
about the fact that the new theory is well defined~\cite{woodard} and stable both {\it outside}
and {\it inside} matter (as an example theories such as $f(R_{\mu\nu})$ or
$f(R_{\mu\nu\rho\sigma})$ involve extra massive spin-2 ghosts~\cite{gh}; we also stress the
Ostrogradksi theorem~\cite{ostro} as recently exposed in Ref.~\cite{woodard}). We can distinguish
two classes,
\begin{enumerate}
\item[3.]\underline{Class C} includes models in which a finite number of new fields are
 introduced. These fields couple to the standard model fields and some of them dominate the matter
 content (at least at late time). This is the case in particular of scalar-tensor theories in
which a
 scalar field couples universally and leads to the class of extended quintessence
 models~\cite{extqu}, chameleon models~\cite{cham} or $f(R)$ models depending on
 the choice of the coupling
 function and potential. For these models, one would have a new sector
 $$
 S_\varphi[\varphi;g_{\mu\nu}]
 $$
 and the couplings of the matter fields will be modified according to
 $$
 S_m[{\rm mat};g_{\mu\nu}]\rightarrow S_m[{\rm mat}_i;A^2_i(\varphi)g_{\mu\nu}].
 $$
 In the case where the coupling is not universal, a signature may be the
 variation of fundamental constants~\cite{uzan02} and a violation of the universality of free fall.
 This was argued to be a general signature of quintessence models~\cite{w2}.
 This is the case in particular in the runaway dilaton model~\cite{runaway}. This class also
 offers the possibility to enjoy $w<-1$~\cite{msu06,boisseau} with a well-defined field
 theory and includes models in which a scalar
 field couples differently to standard matter field and dark matter~\cite{extqudiff}.
\item[4.]\underline{Class D} includes more drastic modifications of gravity with e.g. the
 possibility to have more types of gravitons (massive or not and most probably an
 infinite number of them). This is for instance the case of models involving extra-dimensions such as
 e.g. multi-brane models~\cite{b11}, multigravity~\cite{b12}, brane induced gravity~\cite{b13} or simulated
 gravity~\cite{b14}. In these cases, the new fields modified the gravitational
 interaction on large scale but do not necessarily dominate the matter content
 of the universe. Some of these models may also offer the possibility to mimick an equation of
 state $w<-1$.
\end{enumerate}
These various modifications can indeed be combined to get more exotic models. Since models of
classes C and D involve departure from general relativity, they {\it have to} pass the local tests
on the deviation from general relativity (see Refs.~\cite{msu06,loc2} for examples) both in weak
field in the Solar system and in strong field (timing of pulsars). Both classes lead to a
modification of the Poisson equation on sub-Hubble scales that can be tested using weak lensing or
the large scale structure~\cite{lss}.

Let us stress two important points. First, these models do not address the cosmological constant
problem (see Refs.~\cite{ct,pad1} for reviews on this aspect). Second, at the moment the analysis
of the various cosmological data do not push for any time dependent equation of state and are
completely compatible with the standard $\Lambda$CDM model (see e.g. Ref.~\cite{ct} for a general
review on the constraints, Refs.~\cite{wmap3,data3,data4} for the constraints on a constant
equation of state\footnote{Let us make a short parenthesis concerning models with a constant
equation of state. Such models are often used to evaluate the deviation from a $\Lambda$CDM (see
e.g. Refs.~\cite{wmap3,data3,data4}). Can we determine what kind of models we test with this
assumption? Indeed, using the standard perturbation equations means that we are dealing with class
A models. We can use Eqs.~(\ref{39}-\ref{40}) of section~III.B.2 to characterize the quintessence
models sharing this property. If $w_\de$ is constant then $\rho_\de=\rho_{\de0}(1+z)^{3(1+w_\de)}$
so that, restricting to a flat universe,
$E^2=\Omega_{\mat0}(1+z)^3+\Omega_{\de0}(1+z)^{3(1+w_\de)}$. With the same notations as above, one
gets
$$
\mathcal{V}(z) = \frac{1-w_\de}{2}\Omega_{\de0}(1+z)^{3(1+w_\de)}
$$
and
$$
\mathcal{Q} - \mathcal{Q}_0= \frac{2\sqrt{1+w_\de}}{3w_{\de}}
  \left\lbrace \arg\sinh\left[\sqrt{\frac{\Omega_{\de0}}{\Omega_{\mat0}}(1+z)^{3w_\de}}\right]
  \right.
$$
$$\qquad\qquad\qquad\left.
  - \arg\sinh\left[\sqrt{\frac{\Omega_{\de0}}{\Omega_{\mat0}}}\right]\right\rbrace.
$$
Clearly using a constant equation of state (and larger than $-1$) without modifying the
perturbation equations tests only a very specific class of potential and its meaning in terms of
physical models is far from being clear when one allows $w_\de<-1$.} and Ref.~\cite{setal} for the
case of quintessence models) and there is no need at present for anything else but a cosmological
constant~\cite{six}.

In order to have some handle on the physics behind dark energy, one needs to detail the various
signatures of these models. They may be of different natures: (1) dynamics of the background
(related to the equivalent equation of state for each model), (2) properties of the growth of
structure on sub-Hubble scales, (3) properties of perturbations involving super-Hubble scales,
such as the CMB, (4) non-linear clustering, (5) local tests and (6) strong field effects such as
gravitational waves production and low acceleration regimes (e.g. galaxy rotation
curves\footnote{Even though we do not aim at designing models explaining both the acceleration of
the universe and galaxy rotation curves, the effect of the new fields and in particular the
modification of gravity, if any, has to be taken into account in that regime and may have some
implication for dark matter.}).

From an observational point of view, we can briefly summarize the data sets that allow to probe
these various regimes. (1) The background dynamics can be probed from the Hubble diagram [and more
particularly using SN~Ia; the angular distance-redshift relation from strong lensing; comoving
volume; number counts vs. redshift using galaxy number counts; baryon acoustic oscillation
(BAO)\footnote{The wavelength of the BAO is mainly determined by the wavevector $k=2\pi/s$ with
$$s=\int_{z_{\rm dec}}^\infty c_s(z)[H(z)/H_0]^{-1}\dd z$$ where $c_s=1/\sqrt{3(1+3\rho_{\rm
b}/4\rho_\gamma)}$. Even though it relies on the dynamics of the photon-baryon fluid on sub-Hubble
scales, it would be interesting to check that modification of gravity does not alter this picture.
Let us also emphasize the existence of distortions due to the spacetime shear which enable a
possible test of the Copernician principle using the Alcock-Paczynski test~\cite{aptest}.}; the
CMB shift parameter; big-bang nucleosynthesis; and in the future the determination of luminosity
distance from gravitational waves]. (2) The growth of structures in the Newtonian regime can be
obtained from 3D-galaxy surveys (luminous matter), halo number count vs. redshift using clusters
of galaxies (X-ray, optical/NIR counts and velocity dispersion, Sunyaev-Zel'dovich effect, strong
and weak lensing) and galaxy redshift distribution (dark halos), Lyman-$\alpha$ forest, weak
lensing (2D- survey but with the possible future development of tomography) that gives access to
the matter and gravitational potential power spectra at different redshifts and on different
scales. It is important to separate the linear and non-linear regimes that involves the regime
(4). (3) Super-Hubble perturbations affect mainly the CMB in general and the integrated
Sachs-Wolfe which, in particular when correlated with large scale structures, contains some
information on dark energy. (5) Local tests are usually performed using the parameterized
post-Newtonian formalism in the Solar system and (6) concerns mainly the timing of pulsars and the
emission of gravitational waves.

From a theoretical point of view, it is important to determine which models are in fact two
versions of the same model which means to determine the nature and couplings of the degrees of
freedom (e.g. the fact that $f(R)$ models are nothing else but scalar-tensor theories~\cite{frst}
or that theories involving $f(R,\Box R,\ldots,\Box^nR)$ are $(n+1)$ scalar-tensor
theories~\cite{wands}; the relation between k-essence and quintessence~\cite{cop}) and to which
extent the same set of data can be reproduced by various theories. In this sense the determination
of the number and nature of the new degrees of freedom are important and gives a theoretical
estimate of the complexity of the theory much more accurate than the number of free parameters of
a model. Obviously, a subset of all available data can be explained by models in different classes
(see below) so that our classification will also help quantify the habiliy of a given set of data
to distinguish between theoretical models, which is different from determining which family fits
best.

From a more phenomenological point of view, it is often usefull to rely on an effective
parameterization of the equation of state to describe the change in the dynamics of the expansion.
Although it is a key empirical information on the rough nature of dark energy,  a detailed
description of its properties demands more thoughtful data interpretation. To be useful, the
parameterization has to be realistic, in the sense that it should reproduce predictions of a large
class of models, it has to minimize the number of free parameters, and it has to be related to the
underlying physics (see e.g. Ref.~\cite{lh}). Because the result of a data analysis will
necessarily have some amount of parameterization dependence~\cite{bck}, choosing a specified
physical model strategy seems preferable to break degeneracies. In particular, it enables to
compute without any ambiguity their signature both in low and high redshift surveys, such as the
cosmic microwave background. The increasingly flourishing number of models hampers to provide a
comprehensive set of unambiguous predictions to constrain physical models one by one with
present-day observations, but there are still several benefits in exploring dark energy this way,
in particular when  weak lensing surveys are used together with CMB observations. In this respect
also, our classification may turn to be usefull.

It follows that one is stuck between pragmatism of the data analysis favouring parameterization of
the equation of state (bottom-up approach) and the theorist point of view (top-down approach). It
is indeed clear that the physical nature of dark energy will not directly come out of the
observations and we need obviously to go beyond the measurement of $w$ to have any idea of the
models or classes of models that are likely to be good candidate both to explain the observed
universe and from a theoretical perspective. In this respect, constructing target models in each
classes is a key issue. Both approaches are indeed complementary.

Here, we want to illustrate how perturbations are of importance when one wants to tackle down the
physics behind the acceleration of the universe. In Section~\ref{sec_disc}, we start by recalling
the properties of the background evolution and then we revisit the description of the
perturbations and in particular we focus on their properties that can shed some light on the class
of theoretical models that are best suited (see Fig.~\ref{fig0}). This will lead us to propose a
chain of tests to investigate dark energy. In Section~\ref{sec1}, we consider a first model
inspired by brane world cosmology and then turn, in Section~\ref{sec2}, to the more involved case
of scalar-tensor theories of gravity.

\section{Distinguishing models}\label{sec_disc}

In this section, we compare the information that can be extracted from the background dynamics and
the dynamics of density perturbation at small redshift. This will lead us to propose a chain of
tests to characterize the properties of the dark energy.

\subsection{Background evolution}

Assuming the validity of the Copernician principle, the equation of state of the dark energy is
obtained from the expansion history, assuming the standard Friedmann equation. It is thus given by
the general expression (see e.g. Refs.~\cite{msu06,linder2,linder4})
\begin{equation}
 3\Omega_\de w =-1+\Omega_K + 2q,
\end{equation}
$q$ being the deceleration parameter,
\begin{equation}
 q=-\frac{a\ddot a}{\dot a^2}=-1+\frac{1}{2}(1+z)\frac{\dd\ln H^2}{\dd z}.
\end{equation}
This expression does not assume the validity of general relativity or any theory of gravity but
gives the relation between the dynamics of the expansion and the property of the matter that would
lead to this acceleration if general relativity described gravity. Thus, it reduces to the ratio
of the pressure to energy density only under this assumption. All the background information about
dark energy is encapsulated in the single function $w$.

As long as the background dynamics is concerned, all the observations are related to the function
$E(z)$ defined by
\begin{eqnarray}\label{HH0def}
 E(z)&\equiv&\frac{H(z)}{H_0}.
\end{eqnarray}
In particular, the angular distance is given by
\begin{equation}
 D_A(z)=D_{H_0}\frac{S_K(\chi)}{1+z}
\end{equation}
where $D_{H_0}=c/H_0$. The radial distance $\chi$ is defined by
\begin{equation}
 \chi(z)=\frac{1}{a_0H_0}\int_0^z\frac{\dd z'}{E(z')}
\end{equation}
and  the angular diameter distance $S_K$ is given by
\begin{equation}
 S_K(\chi) = \left\lbrace
 \begin{array}{lc}
   \sin(\sqrt{K}\chi)/\sqrt{K} & K>0 \\
    \chi & K=0 \\
   \sinh(\sqrt{-K}\chi)/\sqrt{-K} & K<0\\
 \end{array}
 \right..
\end{equation}
In the case where the distance duality relation holds, the luminosity distance is given by
\begin{equation}
 D_L(z)=(1+z)^2D_A(z),
\end{equation}
but this may not be the case in models of the class B.

\subsection{Parameterization of the equation of state}

Most data, and in particular supernovae data, are being analyzed using a general parameterization
of the equation of state. These parameterizations are useful to extract model-independent
information from the observations but the interpretation of these parameters and their relation
with the physical models are not always straightforward.

As a first example, recall that general parameterizations of the equation of state as
\begin{equation}\label{eq:para4}
w(a)=w(a_0) + [w(a_m)-w(a_0)]\ \Gamma(a,a_t,\Delta)
\end{equation}
were shown to allow an adequate treatment of a large class of quintessence
models~\cite{para1,para2}. It involves four parameters $\lbrace w(a_0),w(a_m),a_t,\Delta\rbrace$
and a free function $\Gamma$ varying smoothly between one at high redshift to zero today. Even
though it reproduces the equation of state of most quintessence models, it is not economical in
terms of number of parameters since most quintessence potentials involve one or two free
parameters. If one assumes that the parameterization is supposed to describe the dynamics of a
minimally coupled scalar field, the knowledge of $w$ is indeed sufficient but in a more general
case one would need more information. The background dynamics depends on the potential and its
first derivative, which can be related to $w$ and its derivative. Accounting for perturbations,
one needs to know the second derivative of the potential  which can be inferred from $\ddot
w$~\cite{dave}.

Since we expect dark energy to have observable consequences on the dynamics only at late time, one
can consider an equation of state obtained as a Taylor expansion around a pivot point,
\begin{equation}\label{wCPL}
 \widetilde w=w_* + w_a\left(\frac{1}{1+z_*} - \frac{1}{1+z}\right).
\end{equation}
This form depends on only three parameters and is a generalization of the parameterization
proposed by~\cite{para3} and then \cite{para3bis} where $z_*=0$. Whatever the expression for $w$,
it is easy to sort out that
\begin{equation}
 \rho_\de' = -3(1+\widetilde w)\rho_\de
\end{equation}
where a prime denotes a derivative with respect to the number of $e$-folds, $p=\ln(a/a_0)$. It
follows that
\begin{equation}
 \rho_\de=\rho_{\de0}\,\exp\left[{-3\int_0^p(1+\widetilde w)\dd p}\right]
\end{equation}
and thus
\begin{equation}
 \frac{H^2}{H_0^2} = \Omega_{\mat0}\hbox{e}^{-3p} + \Omega_{K0}\hbox{e}^{-2p} +
 \Omega_{\de0}\hbox{e}^{-3\int_0^p(1+\widetilde w)\dd p}.
\end{equation}

Two considerations are in order when using such a parameterization. First, the redshift band on
which this is a good approximation of the equation of state is a priori unknown. Clearly, compared
with the form~(\ref{eq:para4}), it is unlikely to describe dark energy up to recombination time.
Secondly, when combining observables at different redshift such as weak lensing, Sn~Ia and CMB,
one should choose the value of $z_*$ in such a way that the errors in $w_*$ and $w_a$ are
uncorrelated~\cite{hujain} and the pivot redshift is the redshift at which $w$ is best
constrained. In particular, it was argued that it is important to choose $z_*$ for distance-based
measurements. The problem lies in the fact that the pivot redshift is specific to the observable.
In this respect, dark energy models defined by a Lagrangian are more suitable, yielding to a
definite equation of state as a function of redshift, hence more general than a Taylor expansion
around a pivot point. Eventually, one can read out the values of $w_*$ and $w_{a}$ at whatever
redshift. This complication, arising when one wants to combine datasets with different $z_*$, will
also make it more difficult to infer constraints on the physical models from the constraints on
the parameterization (see Refs.~\cite{setal,linder3} for examples).

There is an alternative way to get a first hint on the nature of dark energy. It may be useful to
consider the plane $(w,w')$ where $w'\equiv\dd w/\dd p$. It was recently
shown~\cite{wwprim1,wwprim2} that quintessence models occupy a narrow part of this plane (assuming
$z_*=0$). As discussed later on, the way to relate these parameters to a physical model (that is
the reciproque) is difficult.

In conclusion, a class of equations of states or the $(w,w')$ analysis are usefull tools to
constraint a class of theoretical models independently of the details of each model and the
parameterization must be designed for this class of models. In such a case the position in the
$(w,w')$ can also be related to a particular model in this class. For instance, in the case of
quintessence models where the acceleration is due to the gravitational effect of a slow-rolling
minimally coupled scalar field, the position in the $(w,w')$ plane can be related to the slow-roll
parameters
\begin{equation}
 \varepsilon = \frac{1}{16\pi G}\left(\frac{V'}{V}\right)^2,\qquad
 \eta = \frac{1}{8\pi G}\left(\frac{V''}{V}\right),
\end{equation}
characterizing the shape of the potential $V$ close to the value of the scalar field today since
$w\simeq-1+2\varepsilon/3$ and $w'\simeq-4\varepsilon(\eta-2\varepsilon)/3$. Unfortunately, this
cannot determine the model since the same position can be reached by various models from different
classes.

For background evolution, we thus have various parameterizations that describe class A models and
include the $\Lambda$CDM as a subcase. They enable to quantify, within this particular class, how
close from a $\Lambda$CDM the background dynamics allows to be. In the particular case of
quintessence, the relation with the physical parameters is clear but one should impose a prior on
the parameterizations to ensure that $w\geq-1$ as is the case for these models (see e.g.
Ref.~\cite{setal} for a comparison of a data analysis based on a model and a parameterization).
Indeed, great care is required when interpreting constraints obtained from such a parameterization
when relaxing this prior.

\subsection{Taking perturbations into account}

From the study of the background dynamics, one can, in principle, determine or constrain $w(a)$.
Other sets of data, such as weak lensing and large scale structures, involve density
perturbations. Assuming that the metric of spacetime takes the form
\begin{equation}
 \dd s^2 = -(1+2\Phi)\dd t^2 + (1-2\Psi)a^2\gamma_{ij}\dd x^i\dd x^j,
\end{equation}
where $\gamma_{ij}$ is the metric of the spatial section, the evolution of density perturbations
on sub-Hubble scales is dictated by
\begin{equation}\label{Dpluseq}
 \ddot\delta_\mat + 2H\dot\delta_\mat -\frac{1}{a^2}\Delta\Phi=\mathcal{S}_\de.
\end{equation}
This equation derives from the Euler and conservation equations on sub-Hubble scales, that is from
the conservation of the stress-energy tensor of matter. It could inherit a source term
$\mathcal{S}_\de\not=0$ if the presureless matter is coupled to other matter species (e.g. in
class C).

To close this equation, one needs to use Einstein equations to express $\Phi$. Assuming general
relativity and a $\Lambda$CDM model, the Poisson equation takes its standard form
\begin{equation}
 \Delta\Psi = 4\pi G\rho_\mat a^2 \delta_\mat=\frac{3}{2}\Omega_\mat H^2 a^2\delta_\mat,
\end{equation}
and the two gravitational potentials are related by
\begin{equation}
 \Phi-\Psi = 0
\end{equation}
since the anisotropic stress of radiation is negligible in that regime. When considering dark
energy models, one has to allow possible modifications of these two equations.

First, the second Einstein equation can enjoy a non-vanishing anisotropic stress due to the dark
energy sector so that it takes the general form
\begin{equation}
 \Delta(\Phi-\Psi) = \Pi_\de
\end{equation}
where $\Pi_\de$ can be decomposed as $\Pi_\de = 8\pi GP_\de a^2\Delta\pi_\de$. This term leads to
a general source term $\mathcal{S}_\de + \Pi_\de/a^2$ in Eq.~(\ref{Dpluseq}) but it is worthwile
making the distinction because $\Pi_\de$ alone will be involved in lensing observations that
depend on $\Phi+\Psi$. All classes can induce such a term.

Second, the density perturbation of the dark energy may be non negligible in the Poisson equation
and, as in the case with a modification of gravity, one can expect a modification of
proportionality coefficient that may even be scale dependent. So we will assume that the general
form of the Poisson equation is, in Fourier space,
\begin{equation}
 -k^2\Phi = 4\pi G a^2 F_\mat(k,H)\rho_\mat\delta_\mat+\Delta_\de.
\end{equation}
If $F_m$ depends on $k$ then  the modification will not be degenerate with the normalisation of
the amplitude of the matter power spectrum and can be tested~\cite{bu01} otherwise one will have
to use the growth history to determine whether $F_\mat=1$ or not. Such a term can arise in classes
C (see e.g. Ref.~\cite{sur04}) and D (see Ref.~\cite{km05}). We emphasize that for classes A and
B, the growth factor is scale independent and that the primordial scale-dependence and the growth
factor decouple. This may not be the case in classes C and D when $F_\mat$ is scale dependent.
Thus, it is safer to distinguish $P_\mat(k,z)$ and $P_\Phi(k,z)$.

It follows that in general, the equation of evolution of the growth factor is expected to take the
form
\begin{equation}\label{growtheq}
 \ddot D + 2H\dot D -4\pi G\rho F_\mat D=\frac{\Delta_\de}{a^2}
 +\mathcal{S}_\de.
\end{equation}
Indeed, such an equation is not closed and one needs either to propose a parametrization of the
source term or explicit the evolution equations of the dark sector. Note that while the growth
factor depends on ($F_\mat$, $\Delta_\de$, $\mathcal{S}_\de$), weak lensing involves the
combination $\Delta(\Phi+\Psi)=8\pi G\rho a^2F_\mat\delta_\mat + 2\Delta_\de -\Pi_\de$~\cite{lub}
so that combining them may be very fruitful.

In conclusion, the evolution of perturbations at low redshift of the various classes of models can
be characterized in the Newtonian regime by $(\mathcal{S}_\de,\Delta_\de,F_\mat,\Pi_\de)$. Models
of the class A satisfy $(\mathcal{S}_\de,\Delta_\de,F_\mat,\Pi_\de)=(0,\Delta_\de,1,\Pi_\de)$; for
instance, a $\Lambda$CDM model has
\begin{equation}\label{classLCDM}
(\mathcal{S}_\de,\delta_\de,F_\mat,\Pi_\de)=(0,0,1,0)
\end{equation}
while for quintessence models, $\delta\rho_\de=\dot\varphi\delta\dot{\varphi}
-\dot\varphi^2\Phi+V'\delta\varphi$ so that
\begin{eqnarray}\label{classQ}
(\mathcal{S}_\de,\Delta_\de,F_\mat,\Pi_\de)&\simeq&\left(0,0,1,0\right)
\end{eqnarray}
where $\varphi$ is the quintessence field and $\delta\varphi$ its perturbation. This shows in
particular that quintessence and $\Lambda$CDM can only be distinguished, at low redshift, on the
basis of their equation of state. Models of the class B have the same characteristics than a
$\Lambda$CDM for the density perturbation evolution but have an effective equation of state that
may differ from $w=-1$\footnote{Note that in class B, the effective equation of state derived from
the observed dynamics of the universe is different from the ``true'' equation of state which may
be obtained from the growth factor.}. Models of the classes C and D have more general
$(\mathcal{S}_\de,\Delta_\de,F_\mat,\Pi_\de)$ and in particular $F_\mat\not=1$. We shall give two
examples further on.

\subsection{Summary}

\begin{center}
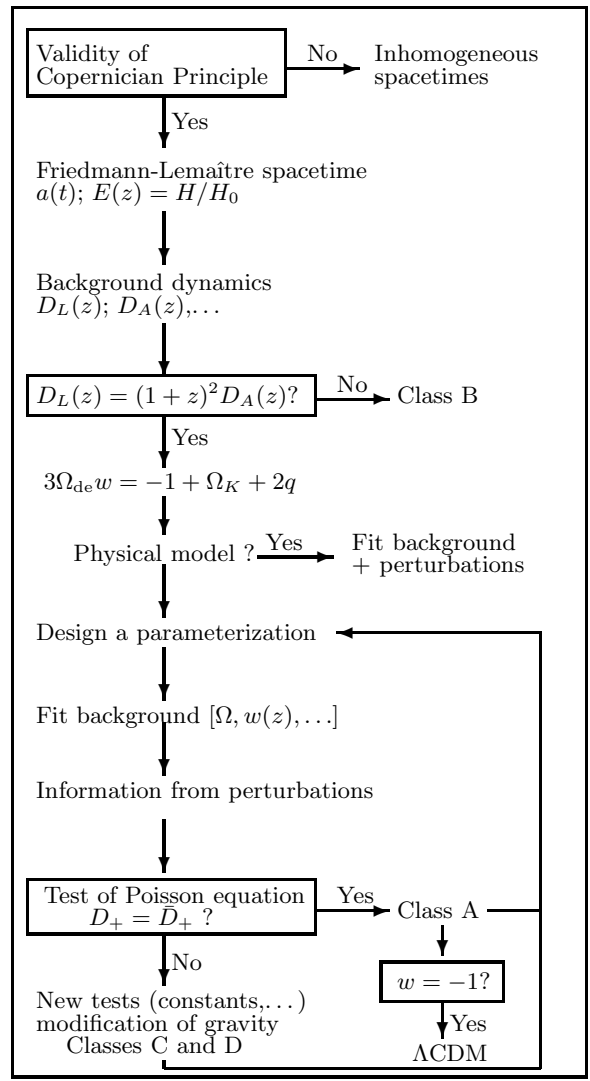
\begin{figure}[ht]
\unitlength=1cm
\begin{picture}(7,14)
 \thicklines
 \put(-.3,.5){\framebox(7.6,14.2){}}
 \put(-.1,13.55){\framebox(3.4,.85){}}
 \put(0,14){Validity of}
 \put(0,13.7){Copernician Principle}
 \put(3.3,13.9){\vector(1,0){1}}
 \put(3.6,14){No}
 \put(4.5,14){Inhomogeneous}
 \put(4.5,13.7){spacetimes}
 \put(1.7,13.55){\vector(0,-1){.7}}
 \put(1.8,13.1){Yes}
 \put(0,12.45){Friedmann-Lema\^{\i}tre spacetime}
 \put(0,12.15){$a(t)$; $E(z)=H/H_0$}
 \put(1.7,12){\vector(0,-1){.7}}
 \put(0,10.95){Background dynamics}
 \put(0,10.65){$D_L(z)$; $D_A(z)$,\ldots}
 \put(1.7,10.5){\vector(0,-1){.7}}
 \put(-.1,9.3){\framebox(3.8,.5){}}
 \put(0,9.45){$D_L(z)=(1+z)^2D_A(z)$?}
 \put(1.7,9.3){\vector(0,-1){.7}}
 \put(1.8,8.9){Yes}
 \put(3.7,9.5){\vector(1,0){1}}
 \put(4,9.6){No}
 \put(4.8,9.45){Class B}
 \put(0.1,8.3){$3\Omega_\de w = -1+\Omega_K + 2q$}
 \put(1.7,8.2){\vector(0,-1){.5}}
 \put(.5,7.35){Physical model ?}
 \put(2.95,7.4){\vector(1,0){1}}
 \put(3.05,7.5){Yes}
 \put(4.2,7.5){Fit background}
 \put(4.2,7.2){+ perturbations}
 \put(1.7,7.3){\vector(0,-1){.7}}
\put(0,6.3){Design a parameterization}
 \put(1.7,6.2){\vector(0,-1){.7}}
 \put(0,5.2){Fit background $[\Omega,w(z),\ldots$]}
 \put(1.7,5.2){\vector(0,-1){.7}}
 \put(0,4.2){Information from perturbations}
 \put(1.7,3.9){\vector(0,-1){.7}}
 \put(-.1,2.4){\framebox(3.8,.7){}}
 \put(0.1,2.8){Test of Poisson equation}
 \put(0.7,2.5){$D_+=\bar D_+$ ?}
 \put(3.7,2.7){\vector(1,0){1}}
 \put(4,2.8){Yes}
 \put(4.8,2.6){Class A}
 \put(1.7,2.4){\vector(0,-1){.7}}
 \put(1.8,1.9){No}
 \put(0,1.4){New tests (constants,\ldots)}
 \put(0,1.1){modification of gravity}
 \put(0.4,.8){Classes C and D}
 \put(5.4,2.5){\vector(0,-1){.4}}
 \put(4.6,1.5){\framebox(1.6,.5){}}
 \put(4.8,1.65){$w=-1$?}
 \put(5.4,1.5){\vector(0,-1){.5}}
 \put(5.5,1.1){Yes}
 \put(5,.7){$\Lambda$CDM}
 \put(6,2.7){\line(1,0){.7}}
 \put(6.7,2.7){\line(0,1){3.7}}
 \put(6.7,6.4){\vector(-1,0){2.7}}
 \put(1.7,0.6){\line(1,0){5}}
 \put(6.7,.6){\line(0,1){3.7}}
 \put(1.7,0.7){\line(0,-1){.1}}
\end{picture}
\caption{A possible chain of tests to unveil the nature of dark energy. The goal is to start from
the more general hypothesis and to incorporate new data and information one by one in order to
check at each step if the hypothesis underlying the equations used in the analysis hold or not.
Here $\bar D_+$ refers to the growth factor predicted from $w(z)$ assuming general relativity. In
particular, it may turn out that a deviation from the Poisson equation may be detected while no
deviation from $w=-1$ is established. This would however require to extend the minimal
$\Lambda$CDM.} \label{figmethod}
\end{figure}
\end{center}

We have argued that the different models appearing in the classification presented in
Fig.~\ref{fig0} can be characterized by a set of functions including the equation of state to
describe the background dynamics and $(\mathcal{S}_\de,\delta_\de,F_\mat,\Pi_\de)$ for the
evolution of the perturbations at low redshift. In most cases, the dependence on the primordial
spectrum and on the dark energy sector decouple in the Newtonian regime. Indeed, more information
would be needed to treat CMB anisotropies and strong field phenomena.

This description can help to get some handle the nature of dark energy and in particular can give
a way to organize the theoretical landscape and the way to distinguish between classes. In
particular, it will be usefull to define both a target model and parameterizations in each class
so that the distance to the $\Lambda$CDM in each category can be quantified. Such
parameterizations exist for the class A but they need to be generalized to other classes and the
constraints on the parameters of these parameterizations for them to describe physical models
actually in these classes should be established.

Let us also mention a difference with a simple fit of models to data. In this case, one ususaly
minimizes the information of the data and the theory together, $I(D,T)$ taking into account a
penalty, $I(T)$, that depends on the number of the free parameters of the theory~\cite{aicbic}.
The total information reduces to $I(D,T)=I(D|T) + I(T)$ with $I(D|T)$ being the opposite of the
likelyhood and various prescriptions for $I(T)$, all depending on the number of free parameters,
exist but in fact it depends on the complexity of the theory. It is in general difficult to define
but it is related to the number, nature and couplings of the new fields of the theory. To this
respect, the $\Lambda$CDM model is the simplest since it involves only 1 new parameter and no new
field. Relying on null test (or exclusion tests) is a way to indicate the degree of complexity
required without having to consider such a statistical issue in the first place (see
Ref.~\cite{doubt} for the implication concerning dark energy).

It follows from the previous discussion that various questions arise in order to grasp some of the
physics behind dark energy. We propose the chart in Fig.~\ref{figmethod} to address them and
outline the dark energy sector. It is based on a series of consistency checks trying to exclude
the most simple models in order to determine how complex a minimal model should be. It also tries
to determine to which extent data can discriminate between the various classes and calls for the
construction of new tests. At the heart of it, is the question of the necessity to extend the
standard $\Lambda$CDM and to determine, in a robust way, how close from it we must be. Indeed
demonstrating the time variabiliy of $w$ is difficult to proove but it may be that other
signatures will be easier to obtain~\cite{dhw}.

\section{First example: DGP cosmology}\label{sec1}

The Dvali-Gabadadze-Porrati (DGP) brane-world model~\cite{b13,dgp} is constructed from a brane
embedded in a 5-dimensional Minkowski bulk. It was shown that because of gravity leakage to the
bulk, it leads to a modification of gravity on large scales and could explain the recent
acceleration phase. The characteristic scale, $r_c$, at which the modification of gravity
manifests itself is related to the two mass scales of the model, the usual 4-dimensional Planck
mass and the 5-dimensional Planck mass, $M_5$. When $M_5\sim 10-100$~MeV, $r_c$ is typically of
the order of $c/H_0$. Interestingly, this model involves only 1 extra-parameter, as the standard
$\Lambda$CDM, but involves new fields.

\subsection{Summary of the cosmological properties}

In the DGP model (see Refs.~\cite{b13,dgp} for a description of the model), the Friedmann equation
takes the form
\begin{equation}\label{fried1}
 H^2 +\frac{K}{a^2} = \left(\sqrt{\frac{8\pi G}{3}\rho + \frac{1}{4r_c^2}}
 +\frac{\varepsilon}{2r_c} \right)^2
\end{equation}
and we consider the $\varepsilon=1$ case. $r_c$ is the new
parameter of the model. We focuse to the low redshift universe
where $\rho$ is dominated by presureless matter. We define
\begin{equation}
 \Omega_\rcto = \frac{1}{4r_c^2H_0^2},
\end{equation}
so that it rewrites as
\begin{eqnarray}\label{HH0}
 E^2(z)\equiv\frac{H^2}{H_0^2}
  = \Theta^2 + \Omega_{K0}(1+z)^2
\end{eqnarray}
where we have defined
\begin{eqnarray}\label{defTheta}
 \Theta(z) \equiv \sqrt{\Omega_\rcto}+\sqrt{\Omega_\rcto+\Omega_{\mat0}(1+z)^3}.
\end{eqnarray}
It follows that we have the constraint
\begin{equation}\label{summ}
 \Theta^2(0)+\Omega_{K0}=\left(\sqrt{\Omega_\rcto}+\sqrt{\Omega_\rcto+\Omega_{\mat0}}\right)^2
 + \Omega_{K0}=1
\end{equation}
the solution of which is
\begin{equation}\label{solsumm}
 \Omega_\rcto=\frac{(-1+\Omega_{\mat0}+\Omega_{K0})^2}{4(1-\Omega_{K0})}.
\end{equation}

By comparing Eq.~(\ref{HH0}) with its standard form, we deduce
that
\begin{equation}\label{rdedgp}
 \frac{8\pi G\rho_\de}{3H_0^2}=
 2\sqrt{\Omega_\rcto}\,\Theta(z),
\end{equation}
or equivalently
\begin{equation}
 \Omega_\de
 =2\sqrt{\Omega_\rc}\left[\sqrt{\Omega_\rc}+\sqrt{\Omega_\rc+\Omega_\mat}\right],
\end{equation}
where
$$
 \Omega_\rc = \Omega_\rcto\frac{H_0^2}{H^2},\quad
 \Omega_\mat = \Omega_{\mat0}(1+z)^3\frac{H_0^2}{H^2}.
$$
In particular, $\Omega_\de+\Omega_\mat+\Omega_K=1$ is equivalent
to the constraint~(\ref{summ}) when evaluated today. Using the
expression~(\ref{HH0}) we obtain that
\begin{eqnarray}\label{dgpeqstate}
 1+w(z)=
 \frac{\Omega_{\mat0}(1+z)^3}{2\Theta(z)\sqrt{\Omega_\rcto+\Omega_{\mat0}(1+z)^3}}
\end{eqnarray}
whatever the curvature. It is easy to sort out that
\begin{equation}
 w'=\frac{\dd w}{\dd p} =-\frac{3}{4}\frac{\sqrt{\Omega_\rcto}\Omega_{\mat 0}(1+z)^3}{
 [\Omega_\rcto+\Omega_{\mat 0}(1+z)^3]^{3/2}}.
\end{equation}

\begin{figure}
 \centerline{\epsfig{figure=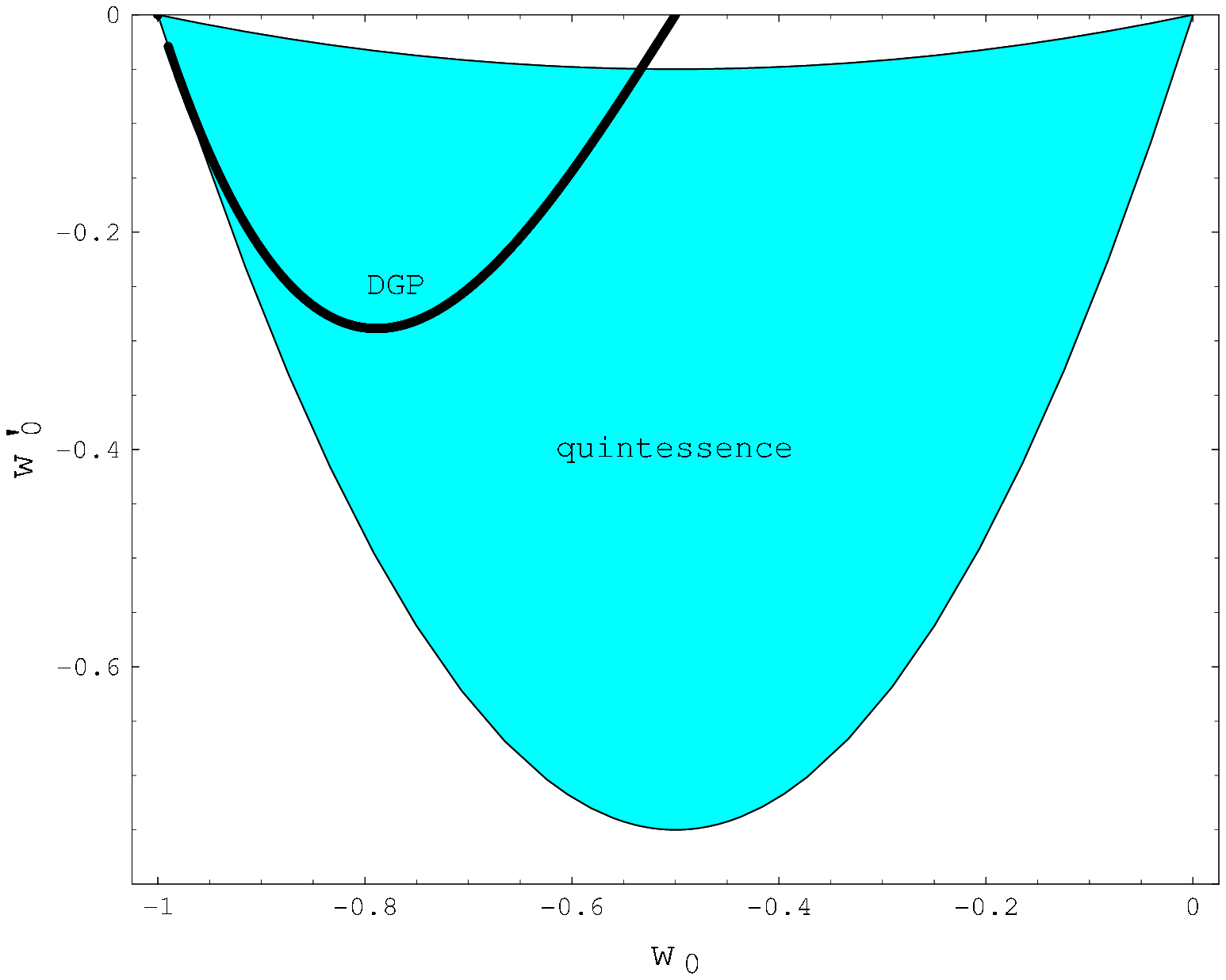,width=7cm}}
 \caption{In the plane $(w,w')$, the filled zone corresponds
 to quintessence models while the very thin black zone corresponds
 to DGP models. We have allowed $\Omega_{K0}$ to vary between $-1$ and $1$ and
 $\Omega_{\mat0}$ between 0 and 1.}
 \label{fig1}
\end{figure}

As a first consequence of this analysis, Fig.~\ref{fig1} shows that today the background evolution
of DGP models can be mimicked by quintessence models. Note that the zone of the plane $(w_0,\dd
w_0/\dd p)$ that corresponds to DGP models remains very thin even when we allow $\Omega_{K0}$ to
vary between $-1$ and $1$ and $\Omega_{\mat0}$ between 0 and 1.

\subsection{Models with equivalent background evolution}

\subsubsection{Fitting the equation of state with a parameterization}

The matching to a given equation of state can be performed in various ways. First, once $z_*$ has
been chosen, we can require to fit $(w,w')$ at the pivot redshift, so that
\begin{equation}\label{fit1}
 w_* = w(z_*),\qquad
 w_a = -(1+z_*)w'(z_*).
\end{equation}
Another route is to require that the equation of state has its correct value today and at a
redshift $z_p$ so that
\begin{equation}\label{fit2}
 w_* = w(z_*),\qquad
 w_a = [w(z_p)-w(z_*)]\frac{(1+z_p)(1+z_*)}{z_p-z_*}.
\end{equation}

\begin{figure}
 \centerline{\epsfig{figure=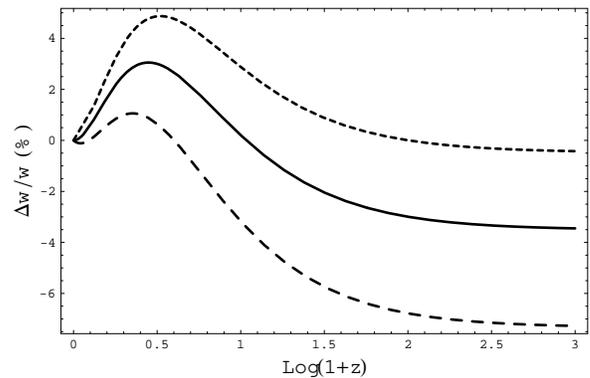,width=8cm}}
 \caption{Comparison of the fits of the DGP equation. The solid line corresponds
 to the fit defined in Eq.~(\ref{fit1}) while the dashed and dotted lines
 correspond to the fit defined in Eq.~(\ref{fit2}) respectively with
 $z_p=3$ and $z_p=1000$. We have assume $\Omega_{\mat0}=0.3$
 and $\Omega_{K0}=0$.}
 \label{fig3}
\end{figure}

In Fig.~\ref{fig3}, we compare these two fits of the DGP equation of state~(\ref{dgpeqstate})
assuming $z_*=0$ and choosing $z_p=3$ and $z_p=100$. By construction both matching lead to the
correct value of the equation of state today ($w_*=w_0$) but the derivative differs by $6.6\,\%$
and $-5.3\,\%$. Indeed these are very small deviations but they may be important while trying to
infer whether a constraint on $(w_0,w'_0)$ points toward a DGP model or a quintessence model. So
the question that arises is how a measured $(w_*,w_a)$ is actually related to the constraint plane
depicted in Fig.~\ref{fig1}, and to the {\it physical parameters} of the model.

\begin{figure}
 \centerline{\epsfig{figure=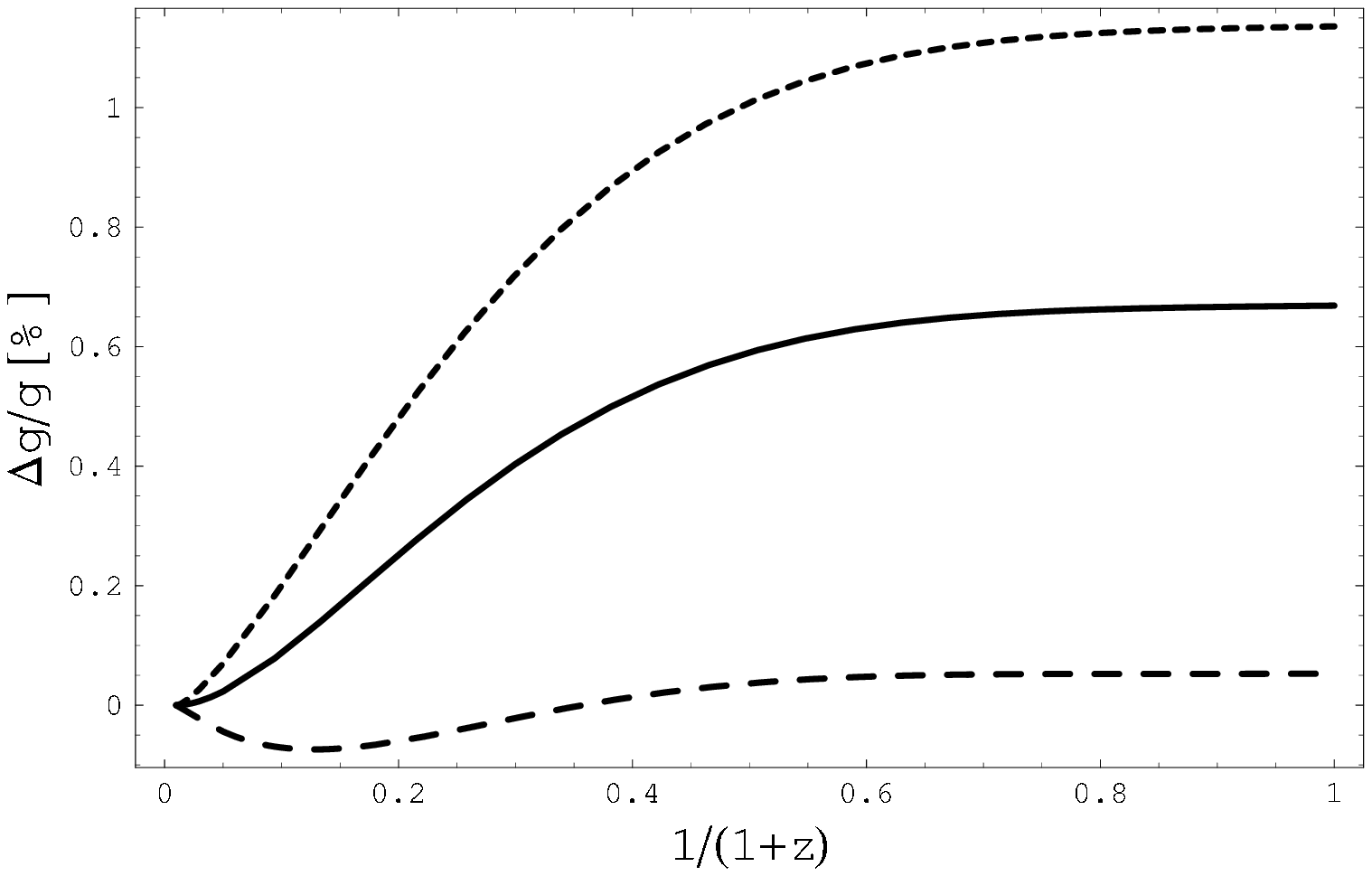,width=8cm}}
 \caption{Comparaison of linear growth factor for the same fits as in Fig.~\ref{fig2} to the
 equivalent quintessence model.}
 \label{fig4}
\end{figure}

Interestingly, as shown in Fig.~\ref{fig4}, the error induced on the growth factor coming from the
error in the fitting of the equation of state is smaller than 1\%. This implies that one cannot
distinguish between the three parametrizations studied in Fig.~\ref{fig3} from the study of
perturbations.

\subsubsection{Equivalent scalar field model}

We can even go a bit further and construct explicitely the quintessence model that reproduces the
DGP dynamics for the background.

Quintessence models reduce to the dynamics of a minimally coupled scalar field, $Q$, evolving in a
potential $V(Q)$. The model is completely specified once the potential is given. It is thus clear
that the energy density is given by
\begin{equation}
 \rho_Q = \frac{1}{2}\dot Q^2 + V
\end{equation}
and the equation of state is
\begin{equation}
 w_Q =\frac{\dot Q^2 -2V}{\dot Q^2+2V}.
\end{equation}
Inverting these two relations, one can reconstruct the potential~\cite{reconstruct} mimicking a
dark energy component\footnote{See also Refs.~\cite{guo} for a similar exercice.} characterized by
$\{\rho_\de(a),w_\de(a)\}$ since we have
\begin{eqnarray}
 \dot Q^2 &=& \rho_\de(a)[1+w_\de(a)],\nonumber\\
 V&=&\frac{1}{2}\rho_\de(a)[1-w_\de(a)].
\end{eqnarray}
It follows that the parametric expression of the potential is
\begin{eqnarray}
 V(z) &=& \frac{1}{2}\rho_\de(z)[1-w_\de(z)],\label{39}\\
 \frac{\dd Q}{\dd z} &=& \pm\frac{\sqrt{\rho_\de(z)[1+w_\de(z)]}}{(1+z)H(z)}.\label{40}
\end{eqnarray}
We can arbitrarily choose the sign of $\dd Q/\dd z$. Using Eq.~(\ref{rdedgp}) and defining
$\mathcal{V}=8\pi GV/3H_0^2$ and $\mathcal{Q}=\sqrt{8\pi G/3}Q$, this system take the form
\begin{eqnarray}
  \mathcal{V}(z) &=& \sqrt{\Omega_\rcto}
  [1-w_\de(z)]\Theta(z),\label{systdgpQ1}\\
 \frac{\dd \mathcal{Q}}{\dd z} &=& \frac{
 \left\lbrace2\sqrt{\Omega_\rcto}[1+w_\de(z)]\Theta(z)\right\rbrace^{1/2}
 }{(1+z)E(z)}.\label{systdgpQ2}
\end{eqnarray}
Once $(\Omega_{\mat0},\Omega_{K0})$ is fixed, $\Omega_\rcto$ is determined and the DGP model we
want to mimick is completely specified. It follows that $w_\de$ is given by Eq.~(\ref{dgpeqstate})
and $E(z)$ by Eq.~(\ref{HH0}) and we have a unique potential obtained from the integration of the
system~(\ref{systdgpQ1}-\ref{systdgpQ2}).

\begin{figure}
 \centerline{\epsfig{figure=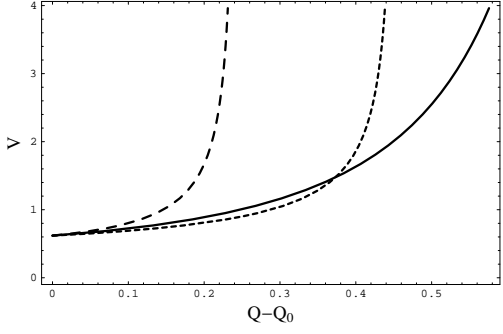,width=7cm}}
 \caption{Reconstruction of the potential of a quintessence model that gives
 the same background evolution than the DGP model when $\Omega_{\mat0}=0.3$
 and $\Omega_{K0}=0$ (solid), $\Omega_{K0}=0.3$ (dashed) and $
 \Omega_{K0}=-0.3$ (dotted).}
 \label{fig2}
\end{figure}

Figure~\ref{fig2} depicts the potentials of the quintessence models that give the same background
dynamics that DGP model for $\Omega_{\mat0}=0.3$ and $\Omega_{K0}=0,\pm0.3$. Indeed, to choose one
or the other model (DGP or quintessence) has some importance and takes some theoretical prejudices
into account.

Note also that, as long as we assume that the growth of density perturbations, is not modified and
is given by a form~(\ref{classQ}), the two models cannot be distinguished.

\subsection{Growth of density perturbation}

The previous result assumed that the growth of density perturbations is dictated by the standard
equation. For the DGP model, this is not the case and two routes have been followed. In the first,
five dimensional effects were neglected~\cite{dgp4d}, which we will refer to as {\it DGP-4D}, so
that
$$
(\mathcal{S}_\de,\Delta_\de,F_\mat,\Pi_\de)=\left(0,0,\frac{2Hr_c}{2Hr_c-1},0\right).
$$
Unfortunately this setting is not compatible with the Bianchi identity. It was recently
argued~\cite{km05} that when five dimensional effects are taken into account we should have
$$
(\mathcal{S}_\de,\Delta_\de,F_\mat,\Pi_\de)=\left(0,0,1+\frac{1}{3\beta},\frac{8\pi
G}{3\beta}\rho_\mat a^2 \delta_\mat\right),
$$
with
\begin{eqnarray}
\beta&=&1-2r_cH\left(1+\frac{\dot H}{3H^2}\right)\nonumber\\
     &=&1-\frac{E}{\sqrt{\Omega_\rcto}}\left[1+\frac{1}{3}(\ln E)'\right].
\end{eqnarray}
We shall refer to this case as {\it DGP-5D}. The influence of this modification of the growth
factor on lensing observables was recently studied in Ref.~\cite{dgp5dlens}.

From these definitions, we deduce that in the DGP-4D case, the equation of evolution of the growth
factor~(\ref{growtheq}) can be rewritten in terms of the growth variable, $g\equiv D/a$, as
\begin{eqnarray}
 g''&+&\left[4+\left(\ln E\right)'\right]g'\nonumber\\
 &+&\left[3+\left(\ln E\right)'-\frac{3}{2}\Omega_\mat
 \frac{E}{E-\sqrt{\Omega_\rcto}}\right]g=0
\end{eqnarray}
where we have used that $2Hr_c=E/\sqrt{\Omega_\rcto}$. In the DGP-5D case, it takes the form
\begin{eqnarray}\label{g5d}
 g''&+&\left[4+\left(\ln E\right)'\right]g'\nonumber\\
 &+&\left[3+\left(\ln E\right)'-\frac{3}{2}\Omega_\mat
 \left(1+\frac{1}{3\beta}\right)\right]g=0.
\end{eqnarray}

\begin{figure}
 \centerline{\epsfig{figure=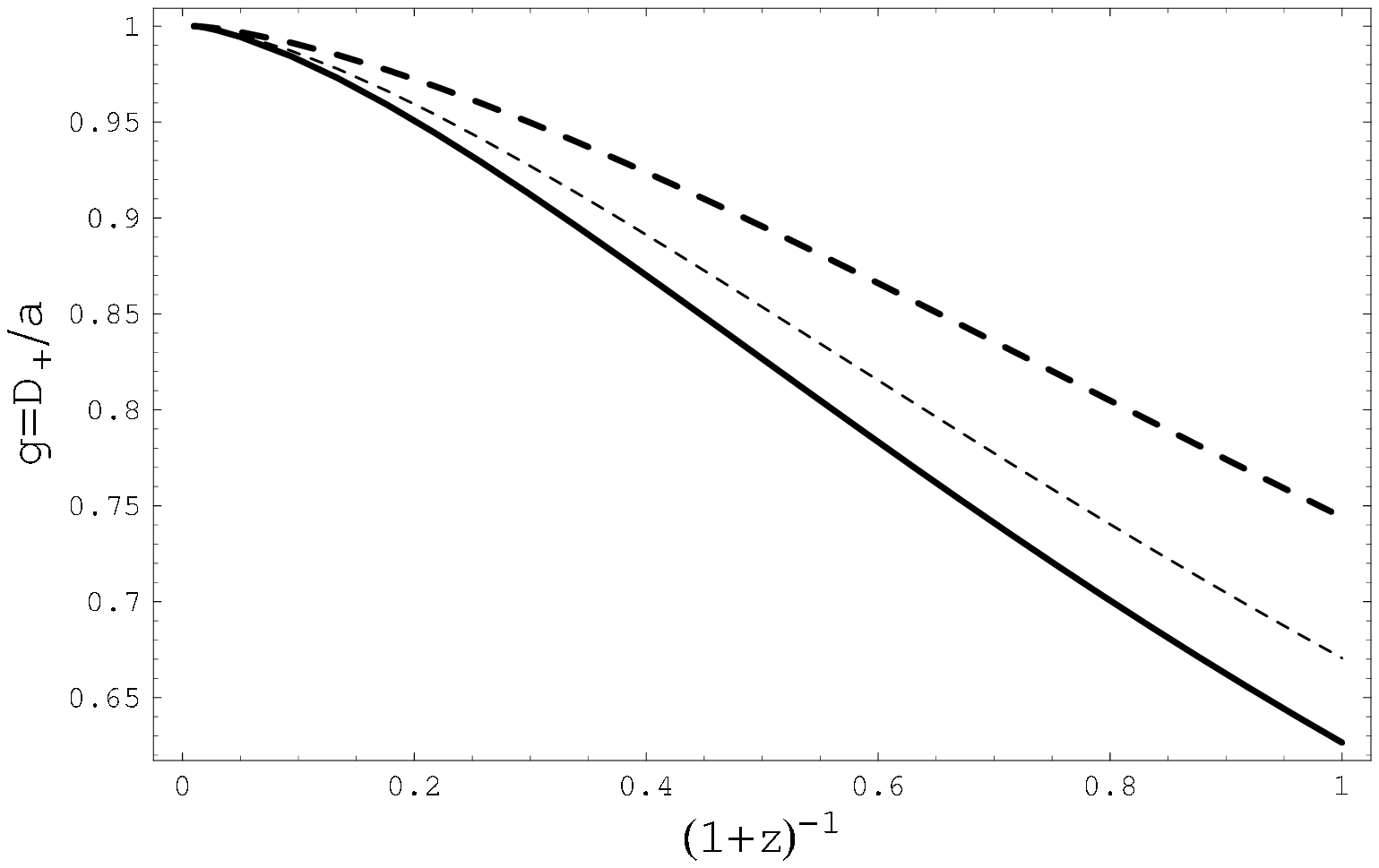,width=8cm}}
 \centerline{\epsfig{figure=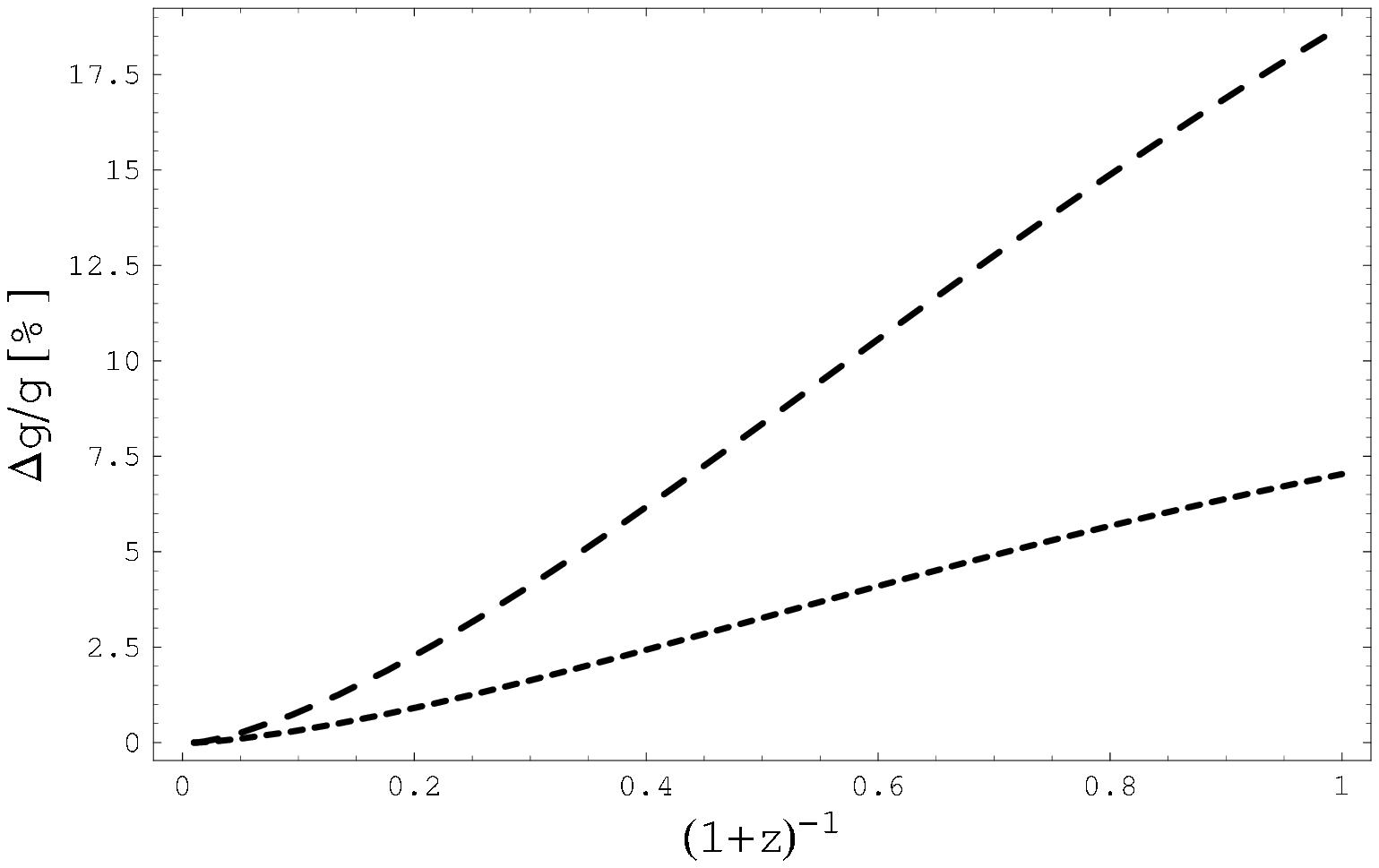,width=8cm}}
 \caption{(Top) Comparison of the linear growth factors of the DGP-5D (solid), DGP-4D (dash)
 and equivalent quintessence (dot) models. (Bottom), the relative error with respect to
 the DGP-5D have been computed and can become large. We have assumed $\Omega_{\mat0}=0.3$
 and $\Omega_{K0}=0$.}
 \label{fig5}
\end{figure}

These two solutions have to be compared to the equivalent quintessence model with same background
dynamics for which
$$
(\mathcal{S}_\de,\Delta_\de,F_\mat,\Pi_\de)\simeq\left(0,0,1,0\right).
$$
Fig.~\ref{fig5} compares the linear growth factor for the DGP-4D, DGP-5D and equivalent
quintessence model. First, we see that the equivalent quintessence model can be discriminated from
the DGP model on the basis of the perturbations (almost 20\% difference at redshift 0). This
clearly illustrates the (trivial) fact that perturbations encode extra-information than the one
given by the bakground dynamics.

We also see that the ``theoretical'' uncertainty on the way to deal with perturbation can lead to
uncertainty of order 5-10\% on the growth factor. It is important to try estimate this
uncertainty, particularly for models that deviates significantly from the standard ones.

\subsection{Summary}

We have considered a DGP model that belongs to the class D. Using the fact that the effective
equation of state is explicitely known we have compared it to a standard parameterization. We have
constructed a quintessence model (class A) sharing exactly the same background dynamics. After
sorting out the form of $(\mathcal{S}_\de,\Delta_\de,F_\mat,\Pi_\de)$, we have shown that
knowledge of the density perturbation can, as expected, help distinguishing these models.

This gives an explicit example in which the background dynamics cannot discriminate between a
class A and class D models. It also shows that one should also try to quantify the ``uncertainty''
of the theoretical models, a difficult task indeed.

\section{Second example: Scalar-tensor theories}\label{sec2}

As a second example, we consider scalar-tensor theories of gravity, which is the simplest example
of a model of class C.

\subsection{Overview}

In  scalar-tensor theories of gravity, gravity is mediated not only by a spin-2 graviton but also
by a spin-0 scalar field that couples universally to matter fields (this ensures the universality
of free fall). In the Jordan frame, the action of the theory takes the form
\begin{eqnarray}\label{actionJF}
  S &=&\int \frac{\dd^4 x }{16\pi G_*}\sqrt{-g}
     \left[F(\varphi)R-g^{\mu\nu}Z(\varphi)\varphi_{,\mu}\varphi_{,\nu}
        - 2U(\varphi)\right]\nonumber\\
        &&   \qquad+ S_m[\psi;g_{\mu\nu}]
\end{eqnarray}
where $G_*$ is the bare gravitational constant from which we define $\kappa_*=8\pi G_*$. This
action involves three arbitrary functions ($F$, $Z$ and $U$) but  only two are physical since
there is still the possibility to redefine the scalar field. $F$ needs to be positive to ensure
that the graviton carries positive energy. $S_m$ is the action of the matter fields that are
coupled minimally to the metric $g_{\mu\nu}$.

It follows that the Friedmann equations in Jordan frame take the form
\begin{eqnarray}
 3F\left(H^2 + \frac{K}{\scalefac^2}\right) &=& 8\pi G_*\rho
  +\frac{1}{2}Z\dot\varphi^2-3H\dot F + U \label{einsteinJF1}\\
 -2F\left(\dot H - \frac{K}{\scalefac^2}\right) &=& 8\pi G_*(\rho+P)
  + Z\dot\varphi^2\nonumber\\
  &&\qquad +\ddot F - H\dot F.\label{einsteinJF2}
\end{eqnarray}
The Klein-Gordon and conservation equations are given by
\begin{eqnarray}
 && Z(\ddot\varphi+3H\dot\varphi)=3F_{\varphi}\left(\dot H + 2H^2
 +\frac{K}{\scalefac^2}\right)\nonumber\\
 &&\qquad\qquad - \frac{1}{2}Z_{\varphi}\dot\varphi^2 -
 U_{\varphi}\label{kleinJF1}\\
 &&\dot\rho+3H(\rho+P) = 0,
\end{eqnarray}
where a subscript $\varphi$ stands for a derivative with respect to the scalar field.

These equations define an effective gravitational constant
\begin{equation}
 G_\eff = G_*/F.
\end{equation}
This constant, however,  does not correspond to the gravitational constant effectively measured in
a Cavendish experiment,
\begin{equation}\label{defgcav}
 G_\cav = \frac{G_*}{F}\left(1+\frac{F_\varphi^2}{2ZF+3F_\varphi^2} \right),
\end{equation}
an expression valid when the scalar field is massless\footnote{As we shall see below, in e.g.
Eq.~(\ref{59}), this is a good approximation on sub-Hubble scales for models where the scalar
field also accounts for the acceleration of the universe. This may not be the case in more
intricate models such as chameleon~\cite{cham}.}. Solar system constraints on the post-Newtonian
parameters imply that $F_{\varphi0}^2/F_0\lesssim 4\times10^{-5}$ so that they do not differ
significantly at low redshift.

\subsection{Sub-Hubble perturbations}

The general cosmological perturbations have been studied in various articles~\cite{sur04,pef} and
we concentrate to the sub-Hubble regime.

The conservation equation of the standard matter are similar to the ones in general relativity so
that $\mathcal{S}_\de=0$. In the Newtonian regime, it can be shown that
\begin{equation}\label{steq1}
 \Psi - \Phi = \frac{F_\varphi}{F} \delta\varphi,
\end{equation}
so that non-minimal coupling induces an extra-contribution to the anisotropic stress,
\begin{equation}
 \Pi_\de = -\frac{F_\varphi}{F} \Delta\delta\varphi.
\end{equation}
The Poisson equation takes the form
\begin{equation}\label{steq2}
 \Delta\Phi = 4\pi \frac{G_*}{F}\rho a^2 \delta_m + \Delta_\de.
\end{equation}
Using the definition~(\ref{defgcav}) of the gravitational constant, we deduce that
\begin{equation}
 F_\mat = \frac{F_0}{F}\left(1 + \frac{F_{,\varphi}^2}{2F + 3F_{\varphi}^2}\right)^{-1}_0
 \simeq \frac{F_0}{F}
\end{equation}
where the last equality has been drawn using the Solar system constraints and should hold at low
redshift (see e.g. Refs.~\cite{sur04,pef} for explicit examples of the redshift variation of this
quantitiy). It can be shown~\cite{sur04} that on sub-Hubble scales,
$\Delta_\de=-(F_\varphi/2F)\Delta\delta\varphi$ so that scalar-tensor models are characterized by
\begin{eqnarray}
(\mathcal{S}_\de,\Delta_\de,F_\mat,\Pi_\de)&\simeq&\left(0,-\frac{F_\varphi}{2F}\Delta\delta\varphi,
 \frac{F_0}{F},-\frac{F_\varphi}{F} \Delta\delta\varphi
\right).\nonumber
\end{eqnarray}
To go further, one needs to determine $\delta\varphi$ and thus use the Klein-Gordon equation which
reduces, in that limit, to
\begin{equation}\label{steq5}
 \left(\Delta - U_{\varphi\varphi}a^2\right)\delta\varphi = F_\varphi\Delta(\Phi-2\Psi)
\end{equation}
from which we deduce
\begin{equation}\label{ddphi}
\left[U_{\varphi\varphi}a^2-\left(1 +2\frac{F_\varphi^2}{F}\right)\Delta
\right]\delta\varphi=F_\varphi\Delta\Phi.
\end{equation}
It follows that $\Pi_\de$ and $\Delta_\de$ are directly proportional to $\Phi$ and the Poisson
equation is given by
\begin{equation}\label{59}
 -k^2\Phi = 4\pi\frac{G_*}{F}\left[1-\frac{k^2F_\varphi^2}{(2ZF+4F_\varphi^2)k^2+2U_{\varphi\varphi}F}
 \right]^{-1}\rho_\mat a^2\delta_\mat.
\end{equation}
When $U_{\varphi\varphi}$ is much smaller than the wavelength of the modes, which a good
approximation in most models such as extended quintessence models, Eq.~(\ref{ddphi}) reduces to
\begin{equation}
 \Delta\Phi \simeq 4\pi G_\cav\rho a^2 \delta_m
\end{equation}
where $G_\cav$ is defined by Eq.~(\ref{defgcav}).

This analysis shows that scalar-tensor models are characterized by
\begin{eqnarray}
(\mathcal{S}_\de,\Delta_\de,F_\mat,\Pi_\de)&\simeq&
\left(0,0,\frac{G_\cav}{{G_{\cav0}}},\frac{F^2_\varphi}{F+2F_\varphi^2}\Delta\Phi\right)
\nonumber\\
&\simeq& \left(0,0,\frac{G_\cav}{{G_{\cav0}}},0\right).
\end{eqnarray}
$F_\mat$ is indeed time dependent but $k$ independent. Two functions of $z$ characterize
scalar-tensor models, the equation of state and $F_\mat$, they are related to the two free
functions of the theory ($F$ and $U$) and influence both the background evolution and the linear
growth of density perturbations.

\subsection{Class C vs class A}\label{secRQ}

As a first exercice, we can quantify the effect of $F_\mat$ and compare a scalar-tensor theory
with a model of class A with the same background dynamics. For that purpose, we need to reconstuct
the scalar-tensor theory, which can be achieved by using th background equations in the
form~\cite{pef}
\begin{widetext}
\begin{eqnarray}
&& \frac{\dd^2 F}{\dd z^2} + \left[\frac{\dd\ln E}{\dd z}-\frac{4}{1+z}\right]\frac{\dd F}{\dd z}+
\left[\frac{6}{(1+z)^2} - \frac{2}{1+z}\frac{\dd\ln E}{\dd z} - 4\frac{\Omega_{K0}}{E^2} \right] F
=
 \frac{2\bar U}{(1+z)^2E^2}+3\frac{1+z}{E^2}\Omega_{\mat0}F_0\label{s1}\\
&&Z\left(\frac{\dd\varphi}{\dd z} \right)^{2} = -\frac{6}{1+z}\frac{\dd F}{\dd z} + \frac{6
F}{(1+z)^2} -\frac{2\bar U}{(1+z)^2E^2} - 6\frac{1+z}{E^2}\Omega_{\mat0}F_0 -
6\frac{F}{E^2}\Omega_{K0}\label{s2}
\end{eqnarray}
\end{widetext}
where $\bar U = UH_0^2$ and $\Omega_{\mat0}=8\pi G_{\cav0}\rho_{\mat0}/3H_0^2\sim8\pi
G_*\rho_{\mat0}/3F_0H_0^2$. As was initially shown in Ref.~\cite{pef}, the knowledge of $H(z)$ and
$D(z)$ allow to reconstruct the two free functions that appear in the microscopic Lagrangian of
the scalar-tensor theory.

The reconstruction can be performed in Jordan frame but it is usefull to shift to Einstein frame
where mathematical consistency is easier to check. To do so, one perform the conformal
transformation $g_{\mu\nu}^*=F(\varphi)g_{\mu\nu}$ so that the potential, $V$, coupling $A$, and
spin-0 degree of freedom, $\varphi_*$, are related to the Jordan frame quantity by
\begin{eqnarray}
 A(\varphi_*) &=& F^{-1/2}(\varphi),\\
 2 V(\varphi_*) &=& F^{-2}(\varphi)U(\varphi),\\
 \left(\frac{\dd\varphi_*}{\dd\varphi}\right)^2&=&\frac{3}{4}
    \left(\frac{\dd\ln F(\varphi)}{\dd\varphi}\right)^2 + \frac{Z(\varphi)}{2F(\varphi)}.
\end{eqnarray}
We usually define $\alpha=\dd\ln A/\dd\varphi_*$ in terms of which the gravitational
constant~(\ref{defgcav}) takes the form $G_\cav=G_*a^2(1+\alpha^2)$. The latter equation can be
rewritten as
\begin{eqnarray}\label{s3}
 \left(\frac{\dd\varphi_*}{\dd z}\right)^2&=&\frac{3}{4}
    \left(\frac{F'}{F}\right)^2 -\frac{F''}{2F} - \left[\frac{1}{2}\frac{\dd\ln E}{\dd z}+\frac{1}{1+z}
    \right]\frac{F'}{F}\nonumber\\
    &&\!\!\!\!\!\!\!\!+ \frac{1}{1+z}\frac{\dd\ln E}{\dd z} - \frac{\Omega_{K0}}{E^2}
    -\frac{3}{2}(1+z)\frac{F_0}{F}\frac{\Omega_{\mat0}}{E^2}
\end{eqnarray}
It follows that $A(\varphi_*)$ and $V(\varphi_*)$ can also be reconstructed parametrically. This
is important because $\varphi_*$ is actually the true spin-0 degree of freedom of the
theory~\cite{def} and must carry positive energy for the theory to be well-defined. This implies
that we must have $(\dd\varphi_*/\dd z)^2>0$. We recall that the spin-2 and spin-0 degrees of
freedom are mixed in Jordan frame so that the positivity of energy does not imply\footnote{If fact
this is the case as soon as $3(\dd\ln F/\dd z)^2>4(\dd\varphi_*/\dd z)^2$ which can happen in
perfectly regular situations and $\varphi$ can become imaginary while $\varphi_*$ remains
well-defined.} that $Z(\dd\varphi/\dd z)^2>0$. In Einstein frame, we have access to $A(\varphi_*)$
and check that it is well defined and to $V(\varphi_*)$. In particular, the sign of
$\dd^2V/\dd\varphi_*^2$ tells us about the sign of the square mass of $\varphi_*$ and it will
indicate the existence of an instability of the model if it were negative.

To estimate the possible magnitude of the effects of this modification of gravity, let us consider
a toy example in which
\begin{equation}\label{paraG}
 F(z) = \frac{1 - \frac{\Delta_G}{2}\tanh\left(\frac{z_G}{\delta z_G}\right)}
  {1 + \frac{\Delta_G}{2}\tanh\left(\frac{z-z_G}{\delta z_G}\right)}.
\end{equation}
With this ansatz, the reconstruction is straightforward since the potential can be obtained
analytically from Eq.~(\ref{s1}) and $\varphi_*(z)$ is deduced by the integration of
Eq.~(\ref{s3}). Eqs.~(\ref{s1}) and~(\ref{defgcav}) can then be used to determine $G_\cav(z)$ that
enters in the equation for the growth of density perturbations. Two questions have then to be
considered. First can such a model be realized by a scalar-tensor theory and second, how different
is the growth factor compared to the one of the class A model sharing the same background
dynamics.

\begin{figure}
 \centerline{\epsfig{figure=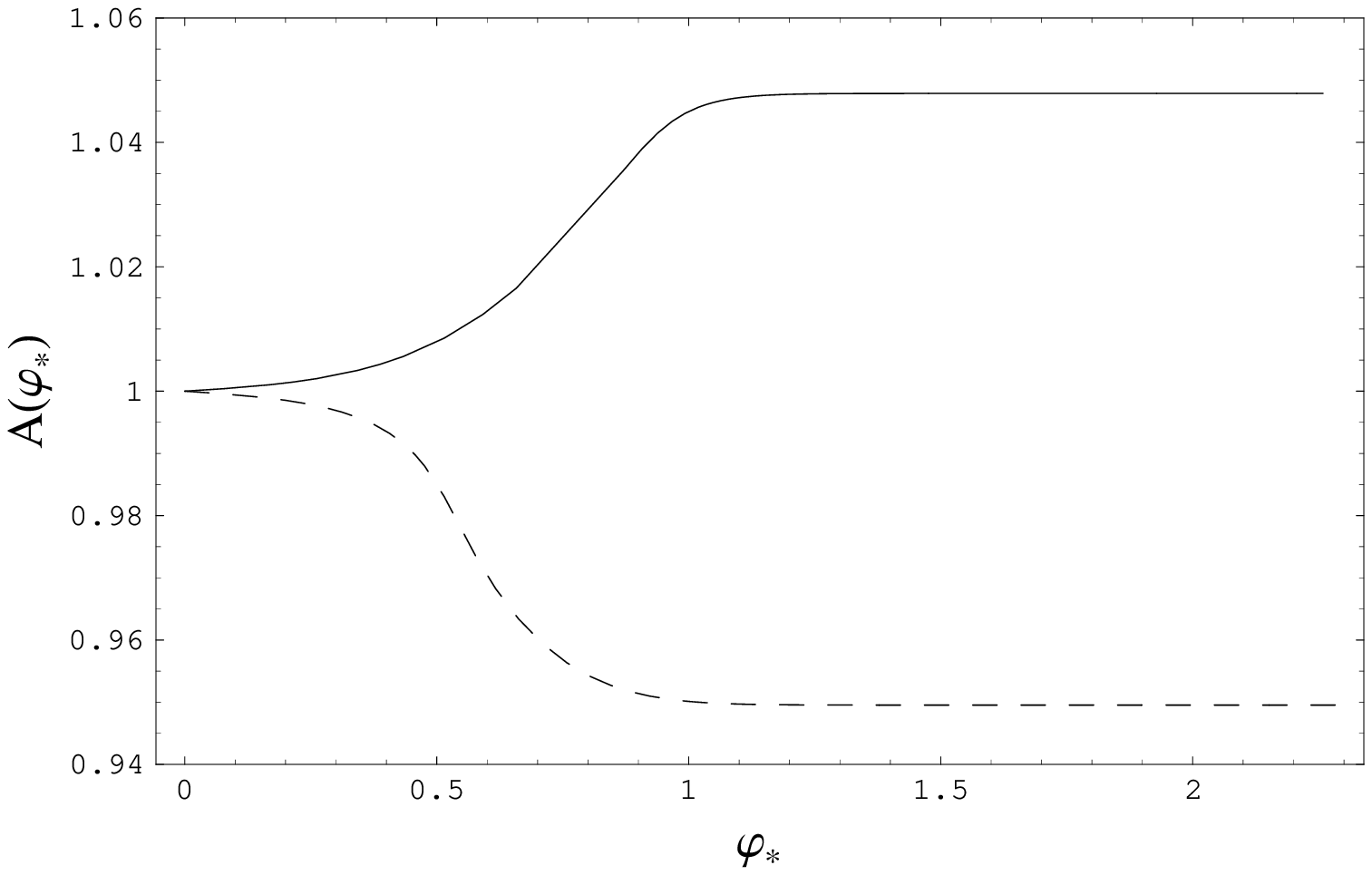,width=7cm}}
 \centerline{\epsfig{figure=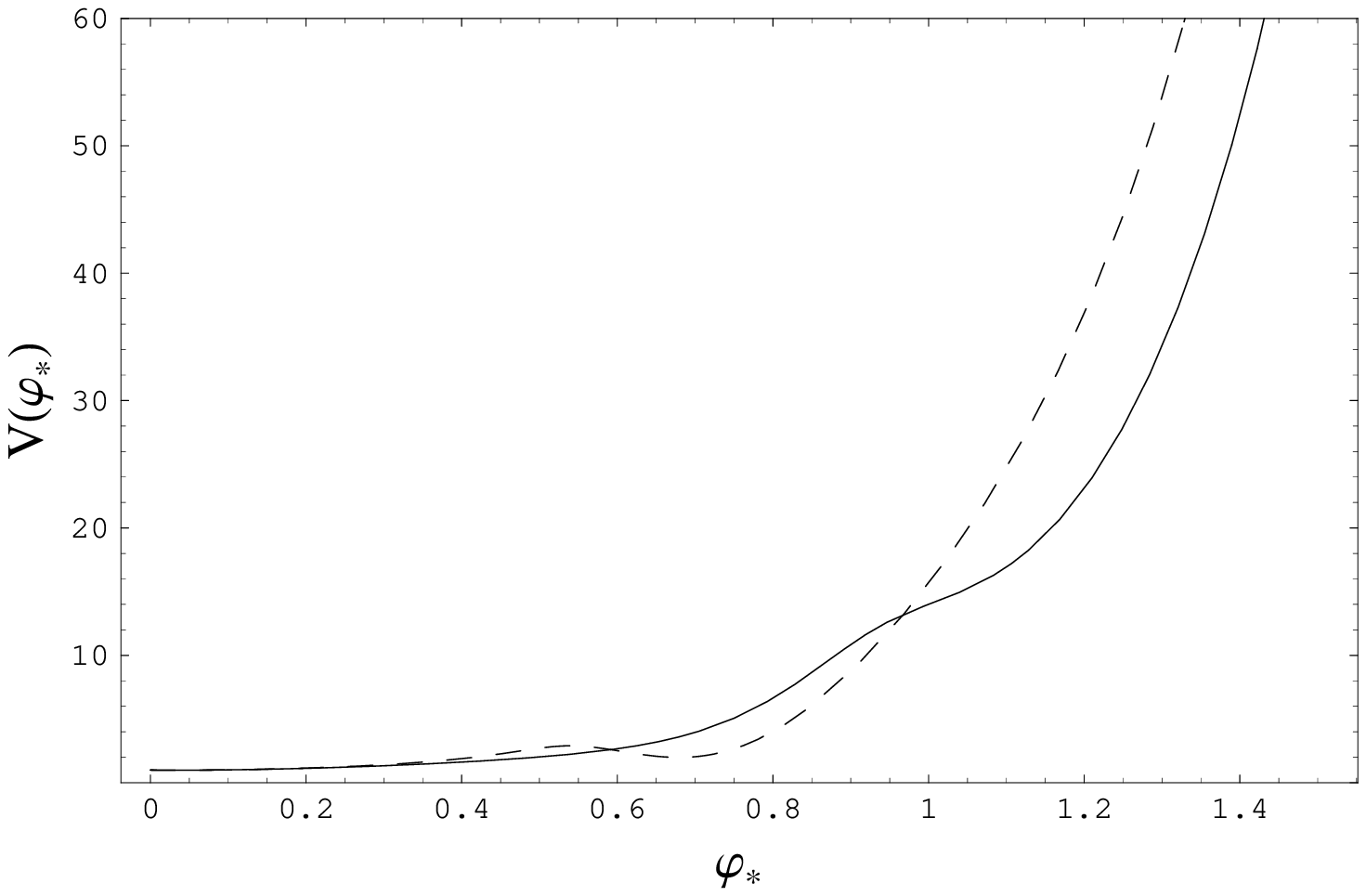,width=7cm}}
 \centerline{\epsfig{figure=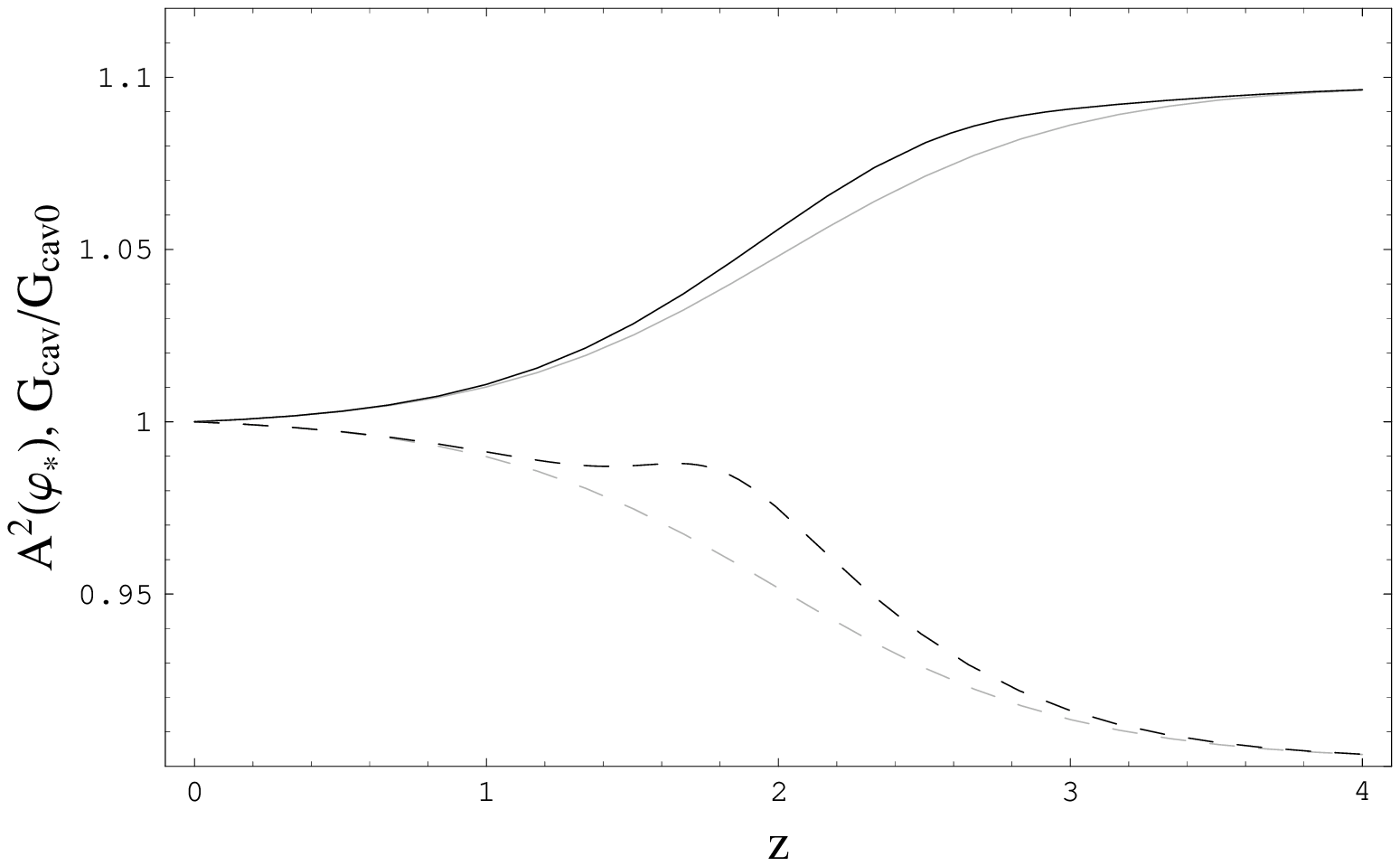,width=7cm}}
 \caption{Reconstruction of the function $A(\varphi_*)$ (top) and $V(\varphi_*)$ (middle)
 of a scalar-tensor theory reproducing an equation of state of the form~(\ref{wCPL})
 with $z_*=0$, $w_*=-0.9$ and $w_a=1.2$ and a gravitational constant of the
 form~(\ref{paraG}) with $z_G=2$, $\delta z_G=1$ and $\Delta_G=0.1$ (solid) or
 $\Delta_G=-0.1$ (dash). We have assumed $\Omega_{\mat0}=0.3$ and $\Omega_{K0}=0$.
 The lower plot compares $G_\cav(z)$ (black) to $A^2(z)$ (light).}
 \label{fig9}
\end{figure}

We consider two examples. In the first one, we assume that $\Delta_G=\pm10\%$ and an equation of
state of the form~(\ref{wCPL}), which will translate in a mild change of $G_\cav$ between low and
hight redshift. Such a modification ensures that we are safe with time variation of the
gravitational constant (see Ref.~\cite{uzan02} for a review and~\cite{cocetal} for constraints
arising from BBN). Fig.~\ref{fig9} gives the reconstructed scalar-tensor theory, that is
$\{A(\varphi_*),V(\varphi_*)\}$, it also compares $A^2(z)$ to $G_\cav(z)$ and shows that they do
not differ significantly. Indeed the potentials and coupling functions may not be easy to justify
from theoretical basis and are completely ad hoc at this point. Note also that not all equations
of state are possible to reconstruct (in most case the scalar-tensor theory is pathologic but the
conclusion mostly depends on the ansatz for $G_\cav(z)$ so that this example is nothing but
illustrative at this point). In the second example, we have played the same game but we have
allowed for a more drastic change in the coupling function, $\Delta_G=0.4$, and we have considered
an equation of state of the form~(\ref{wCPL}) with $w_*=-1.1$ at $z_*=0$ and $w_a=1.6$, which
gives an explicit example of model in which $w_0<-1$. Interestingly, it can be reconstructed to
give a well defined theory (without ghosts) that crosses the so-called phantom barrier. To finish,
we compare the growth factor in these three models with those obtained in the cosmological model
with same background dynamics but which assume general relativity. Not surprizingly, the effect on
the growth factor is typically of the order of $\Delta_G$ (see Fig.~\ref{fig9ter}).

This toy (re)construction tells us that if there are some tensions between background and
perturbation data, so that one is willing to abandon class A models, then a reconstruction will
indicate whether class B models (at least the simple one) are able to lift these tensions and
offer a framework to interpret consistently all data. It is important to keep in mind that one
needs to reconstruct two free functions so that one needs to determine both the expansion history
and the growth of density pertubation observationnaly~\cite{pef}. Otherwise, one must make some
hypothesis on the two free functions~\cite{pef,peri}. It was in particular demonstrated that a
scalar-tensor theory with $U=0$ cannot mimick the background evolution of a
$\Lambda$CDM~\cite{pef} and it is obvious that the scalar-tensor theory corresponding to the
background and growth factor of a $\Lambda$CDM will be defined by $F=1$ and $U ={\rm constant}$.

\begin{figure}
 \centerline{\epsfig{figure=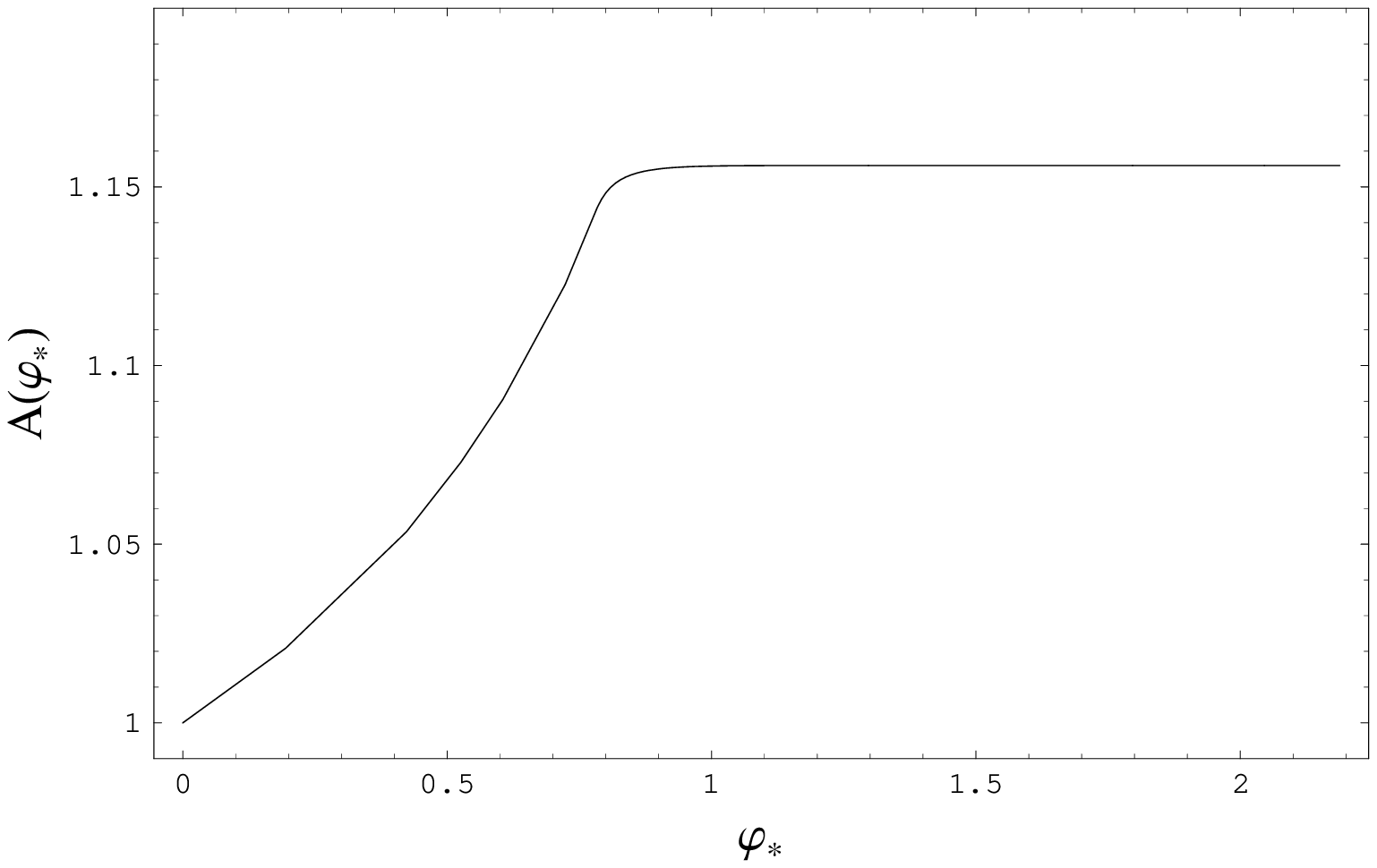,width=7cm}}
 \centerline{\epsfig{figure=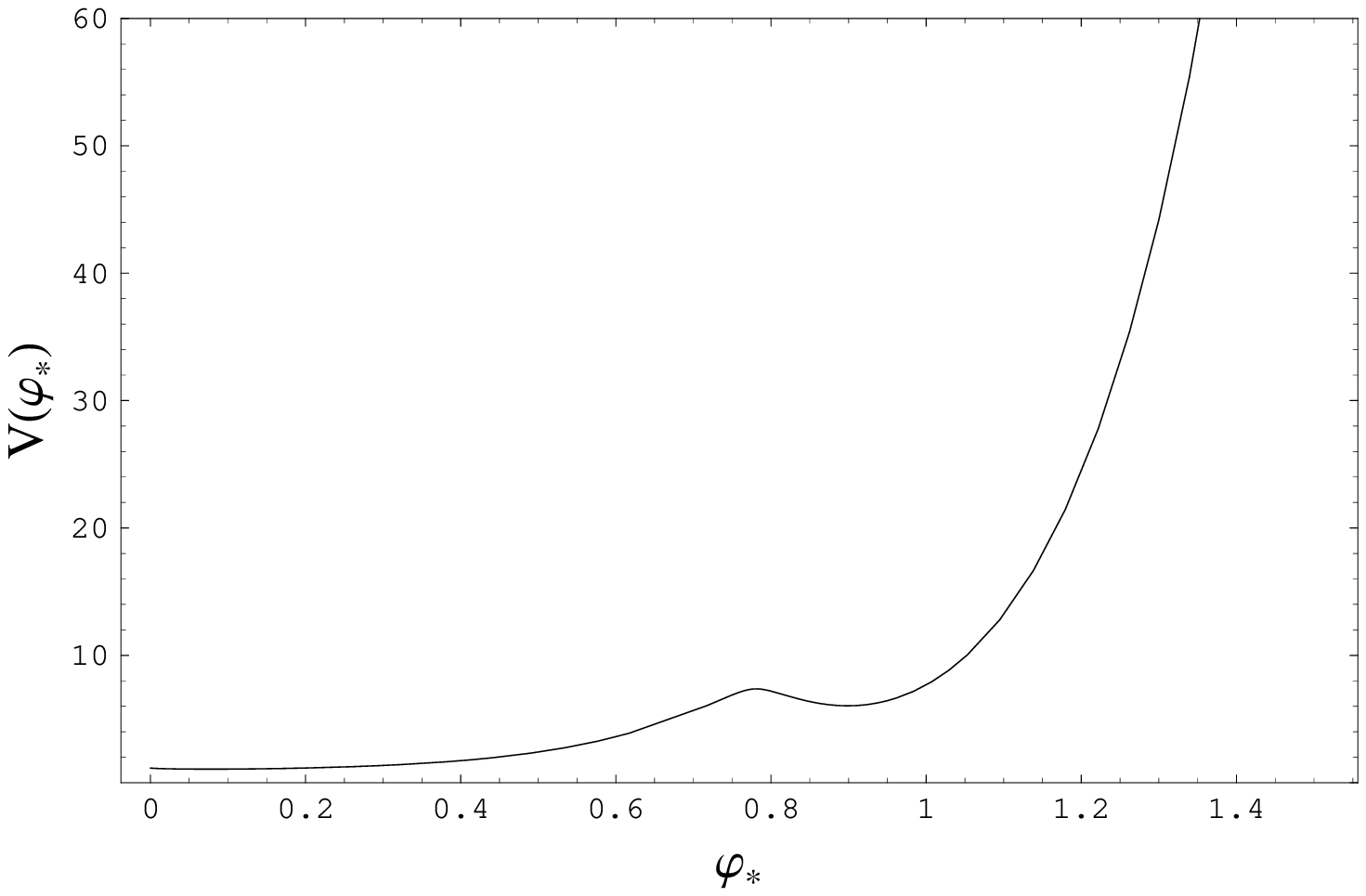,width=7cm}}
 \centerline{\epsfig{figure=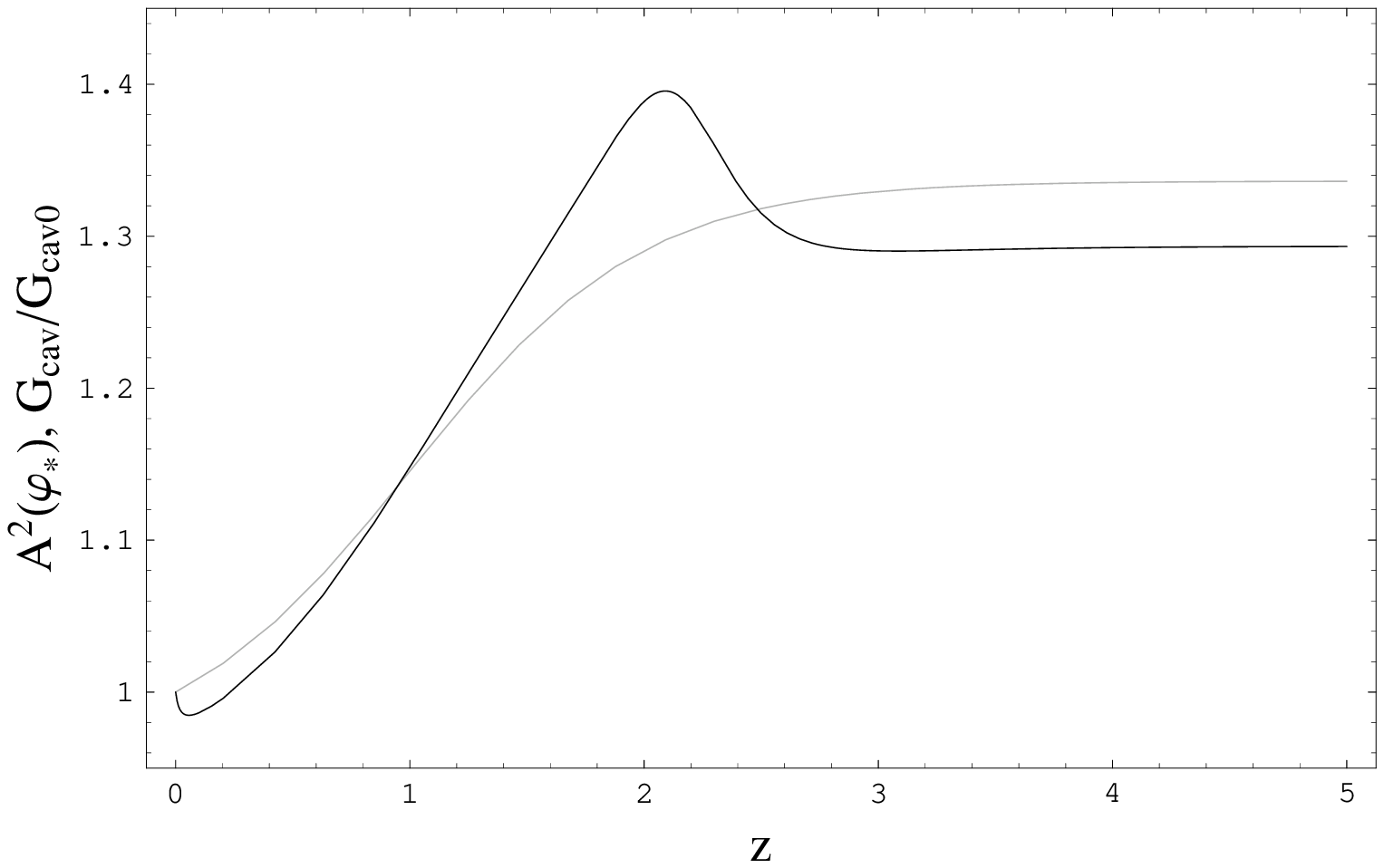,width=7cm}}
 \caption{Reconstruction of the function $A(\varphi_*)$ (top) and $V(\varphi_*)$ (middle)
 of a scalar-tensor theory reproducing an equation of state of the form~(\ref{wCPL})
 with $z_*=0$, $w_*=-1.1$ and $w_a=1.6$ and a gravitational constant of the
 form~(\ref{paraG}) with $z_G=1$, $\delta z_G=1$ and $\Delta_G=0.4$.
 We have assumed $\Omega_{\mat0}=0.3$ and $\Omega_{K0}=0$.
 The lower plot compares $G_\cav(z)$ (black) to $A^2(z)$ (light).}
 \label{fig9bis}
\end{figure}

\begin{figure}
 \centerline{\epsfig{figure=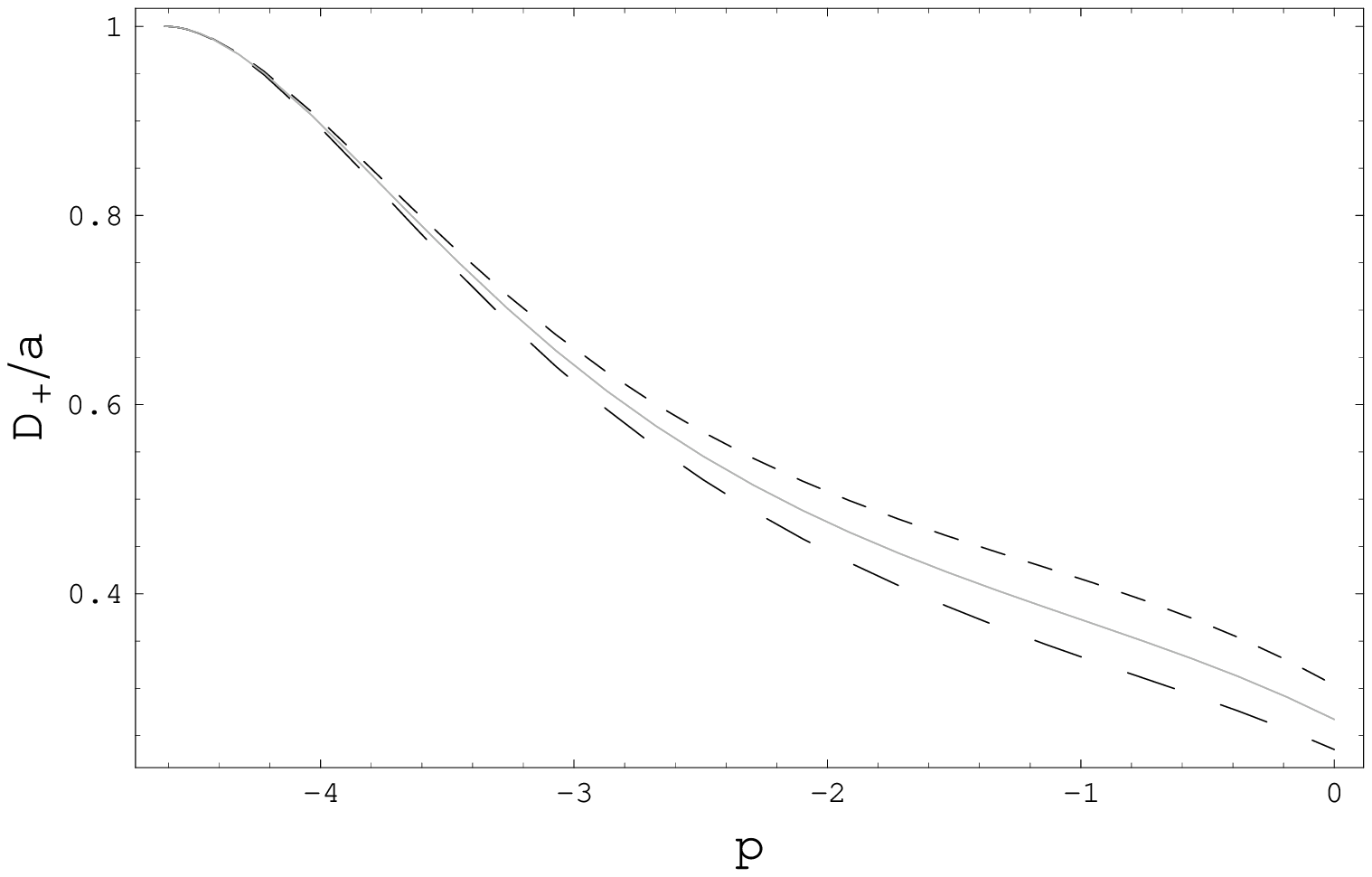,width=7cm}}
 \centerline{\epsfig{figure=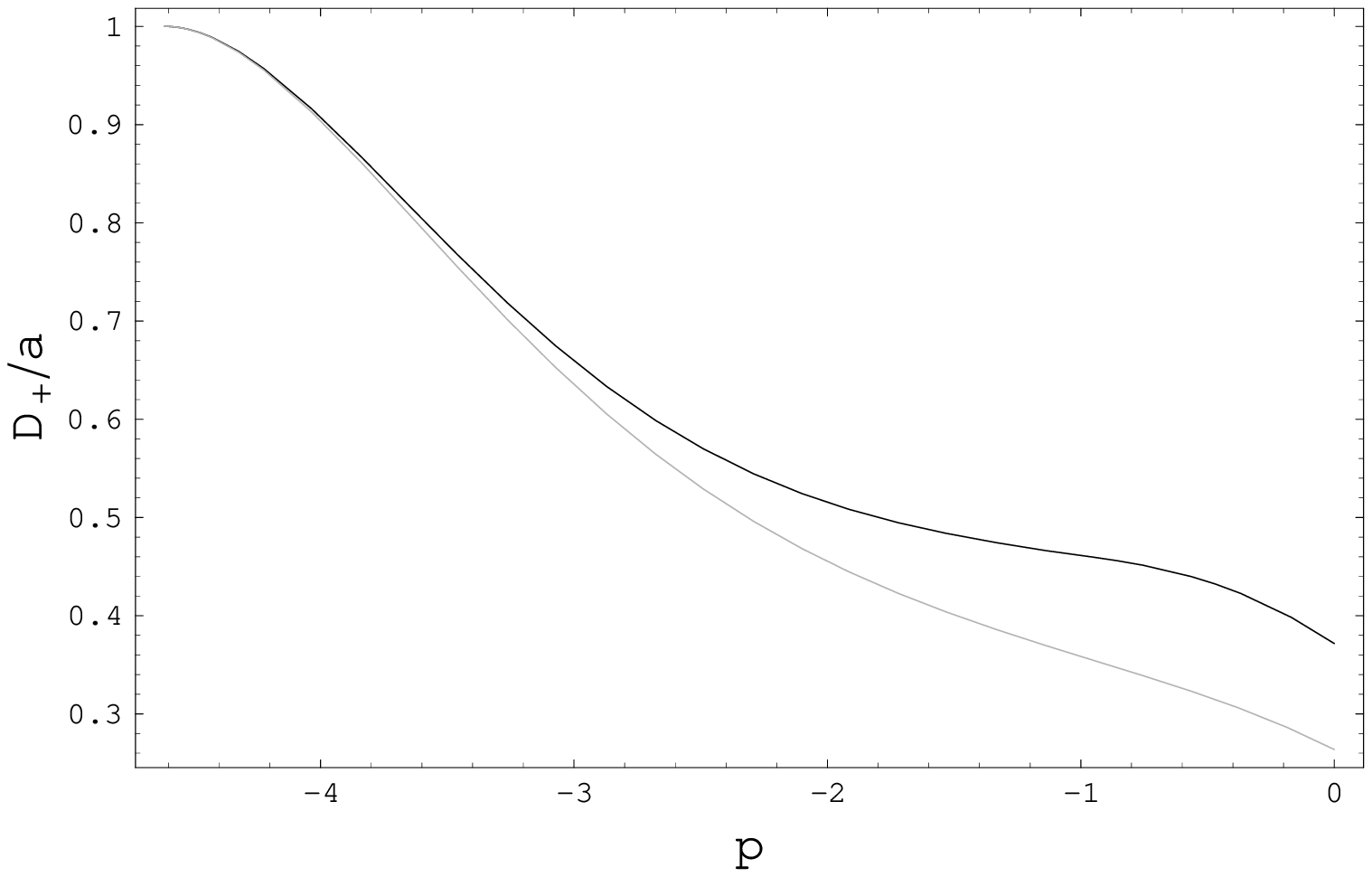,width=7cm}}
 \centerline{\epsfig{figure=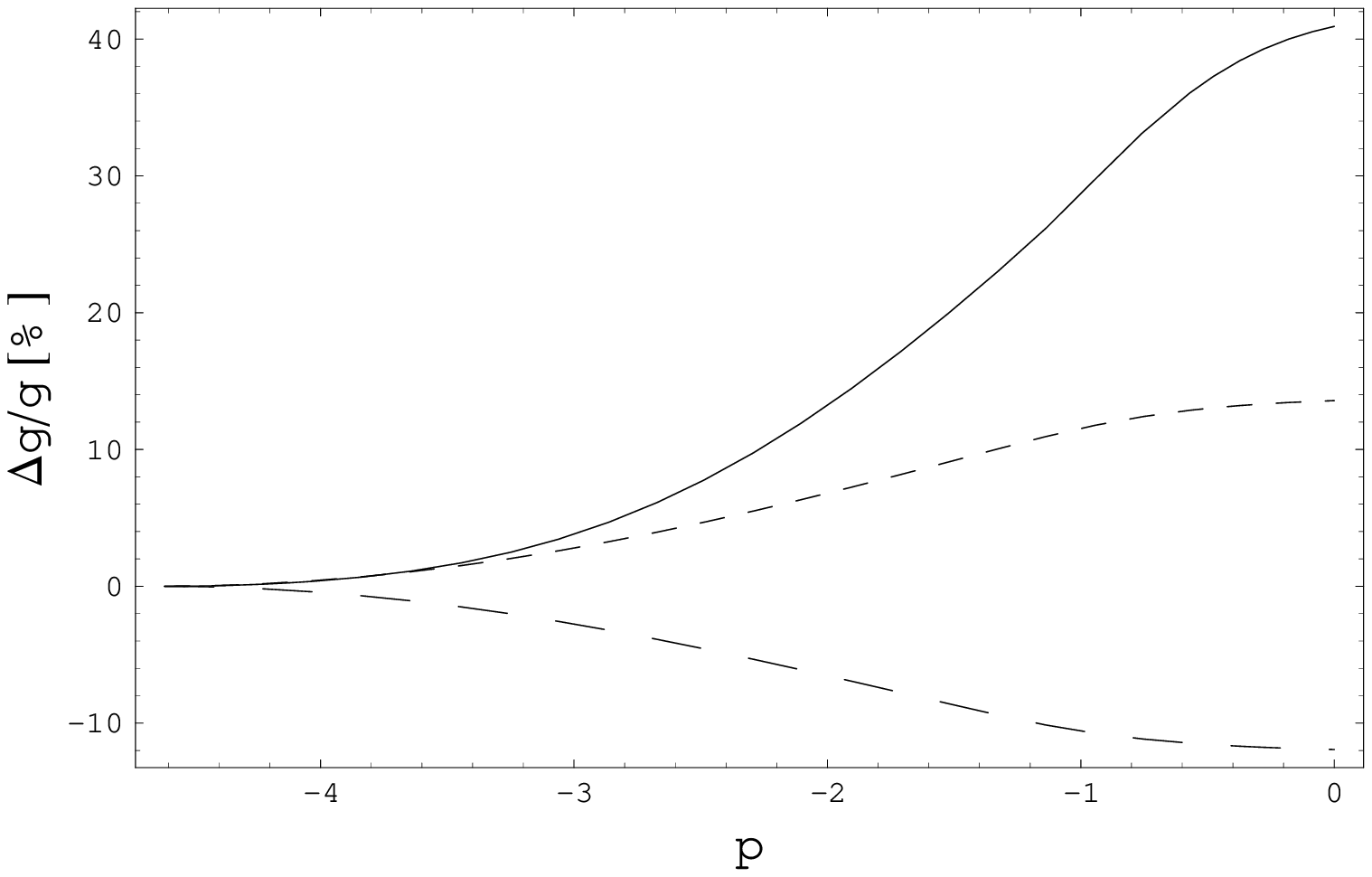,width=7cm}}
 \caption{Comparaison of the growth factor of the scalar-tensor models with the models
 sharing same background dynamics but assuming general relativity. (top) Model depicted in
 Fig.~\ref{fig9} (long dash: $\Delta_G=-10\%$, short dash $\Delta_G=+10\%$, light curve: general
 relativity); (middle) Model depicted in
 Fig.~\ref{fig9bis} and (bottom) relative deviation.}
 \label{fig9ter}
\end{figure}

\subsection{Class C vs class D}

We have  shown that the background dynamics of DGP models can be mimicked by a quintessence model,
but that this was no more true at the perturbation level. It is obvious that a scalar-tensor
theory that can reproduce a given $\{w(z),\beta(z)\}$ will mimick a  DGP model, both for the
background dynamics and perturbation evolution. Such a scalar-tensor theory, if it exists, will be
characterized by
\begin{eqnarray}
 &&E(z)=E_{\rm DGP}(z),\\
 &&\frac{G_{\cav}}{G_{\cav0}} =\frac{1+1/3\beta(z)}{1+1/3\beta(0)}\equiv g_{\rm DGP}(z)\label{d0}.
\end{eqnarray}
In that case, the reconstruction procedure is the following. First, we can eliminate $U$ from
Eq.~(\ref{s2}) to get
\begin{eqnarray}
&&\!\!\!\!\!\!\!\!\!\!\!\!\!\!\!\!Z\left(\frac{\dd\varphi}{\dd z} \right)^{2} =-\frac{\dd^2 F}{\dd
z^2} -\left[\frac{\dd\ln E}{\dd
z}+\frac{2}{1+z} \right] \frac{\dd F}{\dd z}\nonumber\\
&&+2\left[\frac{1}{1+z}\frac{\dd\ln E}{\dd z}- \frac{\Omega_{K0}}{E^2}\right]F
-3\frac{1+z}{E^2}\Omega_{\mat0}F_0.\label{s2bis}
\end{eqnarray}
This equation, together with Eq.~(\ref{defgcav}), yields a non-linear second order differential
equation for $F(z)$ with a source term given by Eq.~(\ref{d0}).

\begin{figure}
 \centerline{\epsfig{figure=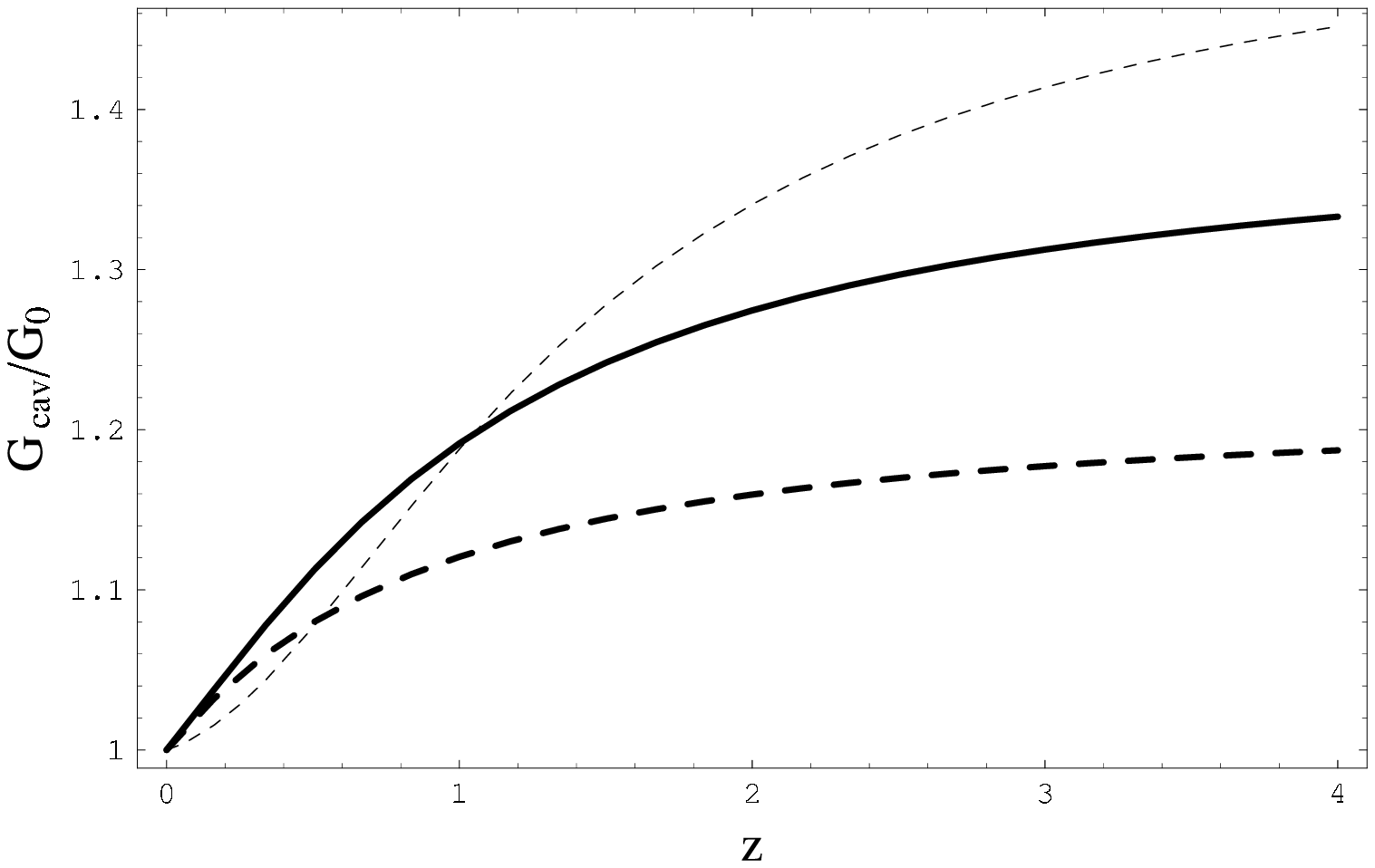,width=7cm}}
 \centerline{\epsfig{figure=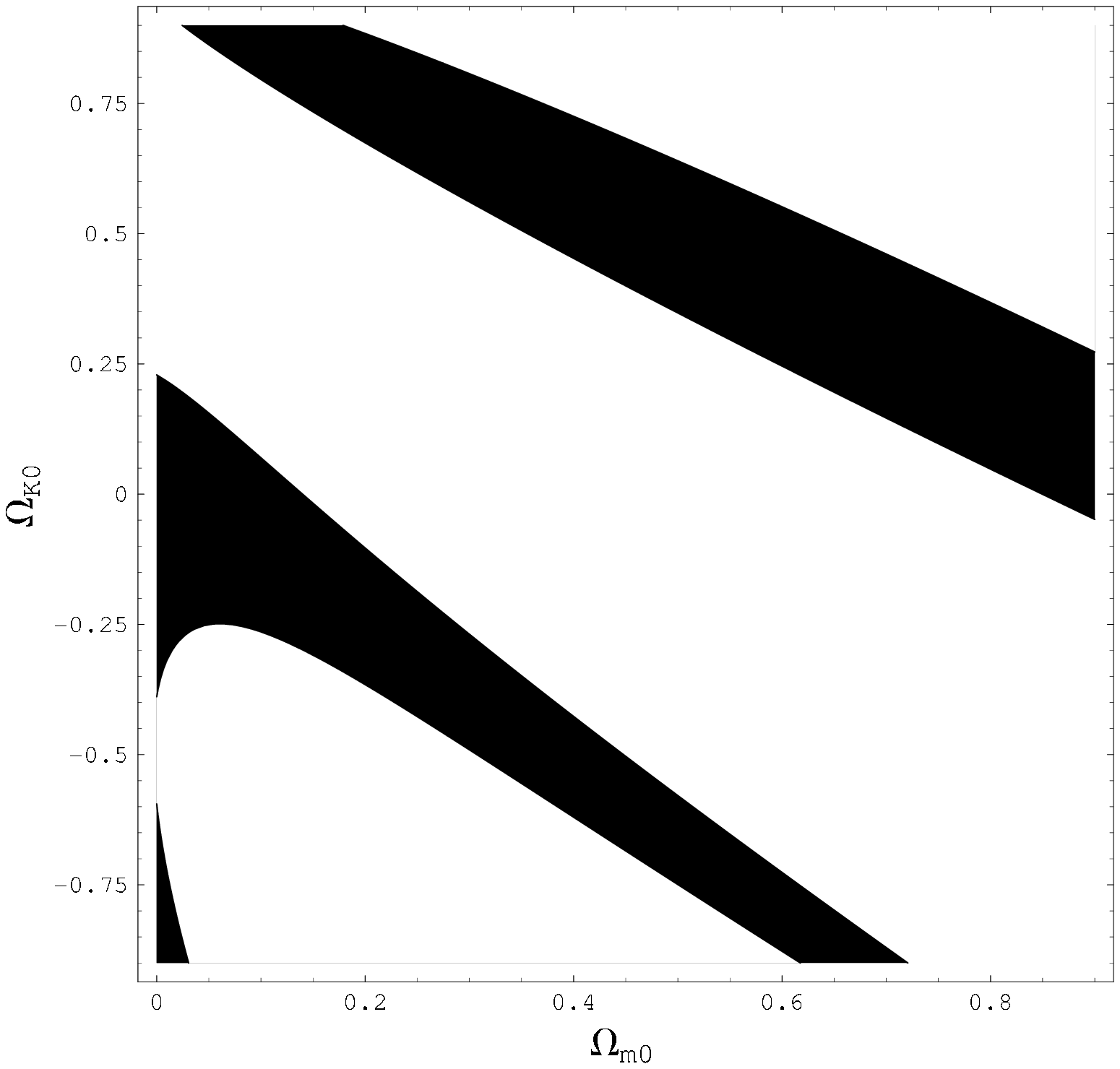,width=7cm}}
 \caption{(top) $G_\cav(z)$ needed for a scalar-tensor theory
 to mimick a DGP model when $\Omega_{\mat0}=0.3$ and $\Omega_{K0}=0$ (solid), $\Omega_{K0}=0.3$ (dashed)
 and $\Omega_{K0}=-0.3$ (dotted).
 (bottom): All DGP models which parameters are lying in the white zone do not fulfill
 the constraint~(\ref{gdgcont}) that will appear on the equivalent scalar-tensor
 theory.}
 \label{fig6}
\end{figure}

Fig.~\ref{fig6} depicts the redshift evolution of the gravitational constant of the equivalent
scalar-tensor theory (if it exists). Let us first stress that the time evolution of $G_\cav$ at
$z=0$ varies a lot with the parameters of the DGP model. The constraint
\begin{equation}\label{gdgcont}
 \left|\frac{\dd\ln G_\cav}{\dd t}\right|_0<6\times10^{-12}\,\rm{yr}^{-1}
\end{equation}
implies that most DGP models, if interpreted as a scalar-tensor theory, are not compatible with
Solar system experiments, since they should fulfill $|\dd\ln G_\cav/\dd
z|_0<5.86\times10^{-2}h^{-1}$. Fig.~\ref{fig6} summarizes the range of cosmological parameters of
the DGP model for which the reconstructed scalar-tensor theory, if it exists, is compatible with
Solar system tests on the time variation of the Newton constant.

The reconstruction is more easily performed by using the Brans-Dicke representation in which
$F=\varphi$ and $Z=\omega(\varphi)/\varphi$ so that
\begin{equation}\label{71}
 G_\cav=\frac{G_*}{\varphi}\frac{2\omega+4}{2\omega+3}.
\end{equation}
This representation is well-behaved in the sense that $\varphi^{\prime2}$ remains positive and the
energy condition reduces to $\omega\geq-3/2$. Using Eq.~(\ref{71}) to express $\omega$ as
\begin{equation}
 2\omega = \frac{4-3\lambda_0g_{\rm DGP}(z)\varphi}{\lambda_0g_{\rm DGP}(z)\varphi-1}
\end{equation}
with $\lambda_0=(2\omega_0+4)/(2\omega_0+3)$, Eq.~(\ref{s2bis}) reduces to closed non-linear
second order equation for $\varphi$. To performed the reconstruction, we need to determine
$\varphi_0$, $\varphi'_0$ and $\omega_0$. While it is always possible to set $\varphi_0=1$, we
need to determine the two others. This can be done by using the expressions~\cite{def,will,pef} of
the PPN parameters
\begin{eqnarray}
 \gamma^\ppn-1 &=& -\frac{1}{\omega_0+2},\\
 \beta^\ppn -1 &=&\frac{1}{4}\frac{1}{(2\omega_0+3)(\omega_0+2)^2}\frac{\omega_0'}{\varphi_0'}
\end{eqnarray}
and the expression of the time derivative of $G_\cav$ to get
\begin{eqnarray}
 \varphi'_0 &=& -\left(\frac{G'_\cav}{G_\cav}\right)_0\left(1-4\frac{\beta^\ppn -1}{\gamma^\ppn-1}
 \right)^{-1}\\
 \omega_0 &=& -\left(\frac{1}{\gamma^\ppn-1}+2 \right).
\end{eqnarray}
We see that the question of the reconstruction of the DGP cosmological dynamics by a scalar-tensor
theory depends on the prediction of the DGP theory in the Solar system, an issue still not
settled\footnote{In Eq.~(\ref{defTheta}) and following describing the background dynamics of DGP,
$\Omega_{\mat0}$ has been defined as $\Omega_{\mat0}=8\pi G\rho_{\mat0}/3H_0^2$ without
questioning the fact that $G$ is the Newton constant, the numerical value of which is determined
by a Cavendish-like experiment today [see Eq.~(\ref{defgcav}) for the similar issue in
scalar-tensor theory]. Without more information, we have to consider it as a pure bare parameter.
The Poisson equation for DGP-5D takes the form
$$
\Delta\Psi
=\frac{3}{2}\bar\Omega_{\mat0}H_0^2\frac{G}{G_{\cav0}}\left(1+\frac{1}{3\beta}\right)\frac{\delta_\mat}{a}
$$
with $\bar\Omega_{\mat0}=8\pi G_{\cav0}\rho_{\mat0}/3H_0^2$. It follows that we can conjecture
(see also Ref.~\cite{lue}) that $G_\cav=G(1+1/3\beta)$ so that
$$
F_\mat=\frac{G_\cav}{G_{\cav0}}=
\left(1+\frac{1}{3\beta}\right)/\left(1+\frac{1}{3\beta_0}\right).$$ This points out a difficulty.
When we compare the background dynamics of DGP to another theory without having a full
understanding of the weak field limit, we cannot be sure that what we define as $\Omega_{\mat0}$
in each theory refers to the same quantity. Fortunately this does not change the conclusions of
\S~III.A.2 but can affect those of this section. Note also that it was shown~\cite{lue} that in
the Solar system the gravitational constant was shifting from $G$ to $4G/3$ when moving away from
the Sun. While the effect of DGP on light bending ans periheli precession have been worked out,
the interpretation of the PPN parameters in this framewok is not settled yet. The same issue needs
to be studied in any model of the classes C and D.}~\cite{lue}.

\begin{table*}
\caption{Summary of the properties of the models considered here. We emphasize the predictions
that one computes in these models (Y: yes, N: no, controversial: there is no current agreement at
the moment).} \label{tab:1} \centering
\begin{tabular}{c|c|ccccccc}
 \hline
  model & class & background & Newtonian & cosmological& $\quad$non-linear$\quad$
  & Solar  & strong & $w<-1$  \\
     &  &  & $\quad$perturbations $\quad$& perturbations & regime
  & system& field & $\quad$possible \\
  \hline\hline
  Quintessence & A & Y & Y & Y & almost & Y & Y & N \\
  Scalar-tensor & C & Y & Y & Y & N & Y & Y & Y \\
  DGP & D & Y & Y  & controversial & N & controversial & Y & Y \\
  \hline
  \end{tabular}
 \end{table*}

\begin{table*}
\caption{Summary of the comparisons performed in our exercices and of the possibility of two
models to share the same predictions. (bgd=background; Newt. pert.= density perturbation in the
Newtonian regime).} \label{tab:2} \centering
\begin{tabular}{c|ccc}
 \hline
       & $\qquad$bgd$\qquad$ & bgd + Newt. pert.$\qquad$ & bgd + Newt. pert. + Solar syst. \\
  \hline\hline
  DGP vs quintessence & Y & N & N  \\
  DGP vs scalar-tensor & Y & ? & N \\
  \hline
  \end{tabular}
 \end{table*}

Note that in Ref.~\cite{km05}, it was suggested that Eq.~(\ref{g5d}) implies that
$\omega(z)=3[\beta(z)-1]/2$. Comparing with Eq.~(\ref{71}), this would mean that $G_\cav(0)$ is
identified with $G_*/F_0$, which is the case only if $\omega_0\gg1$ and that the redshift
dependence of $\varphi$ is neglected. In such a case, one can check that $\omega<-3/2$ so that
there is no hope to reconstruct a well-defined scalar-tensor theory. This may reflect the fact
that the DGP model contains ghost modes around the self accelerating solution~\cite{gmode}.

Playing with the parameters $\gamma^\ppn$ and $\beta^\ppn$, we can reconstruct well-defined
scalar-tensor theories for some values. Indeed, there is no insurance at this stage that it really
describes a DGP model. Anyway, it would mean that a DGP model and the scalar-tensor could share
the same background and perturbation evolution. Of course, this does not mean that DGP-5D are
scalar-tensor theories which would require a complete mapping of the degrees of freedom of each
theory. Also, the two theories can be hoped to be discriminated by other features that are not
discussed here such as the high-redshift properties (CMB, BBN), primordial predictions, strong
field effects (black holes and gravitational waves emission) and, as discussed above and
illustrated on Fig.~\ref{fig6}, local test of gravity. While all these predictions are available
for scalar-tensor theories, this is not yet the case for DGP models (see table~\ref{tab:1}).

\section{Conclusions}

In this article, we have discussed various physical models of the dark energy sector that can be
the cause of the acceleration of the universe, following a physically oriented classification
relying on the nature and coupling of the degrees of freedom of dark energy as starting point. For
each of the four classes, we have described specific signatures and tests (see Fig.~\ref{fig0}).

Beside the evolution of the background, we insisted on the central role of the perturbations to
distinguish between the four classes of models. Restricting to the Newtonian limit, we propose to
characterize them by a new set of functions besides the equation of state. According to this
classification scheme, the determination of the nature of dark energy can be outlined following a
series of consistency checks in order to progressively exclude classes of models. Also, it can be
used to quantify the departure from a pure $\Lambda$CDM in each class and to define target models
in each class. Maybe it will just let us with the initial cosmological constant problem to
face~\cite{pad1} after alternatives offered in the context of (well-defined) field theories have
been exhausted. Let us recall that this analysis let a large class of solutions unexplored and
also calls for tests of the Copernician principle.

As exercices, we have constructed models from different classes that share the same background
dynamics and eventually the same growth history of density fluctuations (see table~\ref{tab:2} for
a summary). These models may be differentiated on the basis of local experiments, strong fields
effects and of large scale perturbations (including CMB). This illustrates the importance of the
theoretical prejudices when parameterizing the dark energy and the complementarity of data sets.\\

\section*{Acknowledgments}

It is a pleasure to thank Nabila Aghanim, Francis Bernardeau, Luc Blanchet, C\'edric Deffayet,
Gilles Esposito-Far\`ese, Cyril Pitrou, Simon Prunet, Yannick Mellier, Carlo Schimd, and Ismael
Tereno for continuing discussions on this topic and others. Part of these thoughts were motivated
by the exciting discussions of the Cape Town cosmology meeting (July 2005) as well as those
stimulated by the ANR project ``Modifications \`a grande distance de l'interaction
gravitationnelle, th\'eorie et tests observationnels''.


\end{document}